\input harvmac
\input psfig
\input epsf
\noblackbox

\newcount\figno
 \figno=0
 \def\fig#1#2#3{
\par\begingroup\parindent=0pt\leftskip=1cm\rightskip=1cm\parindent=0pt
 \baselineskip=11pt
 \global\advance\figno by 1
 \midinsert
 \epsfxsize=#3
 \centerline{\epsfbox{#2}}
 \vskip 12pt
 {\bf Fig.\ \the\figno: } #1\par
 \endinsert\endgroup\par
 }
 \def\figlabel#1{\xdef#1{\the\figno}}

\def\encadremath#1{\vbox{\hrule\hbox{\vrule\kern8pt\vbox{\kern8pt
 \hbox{$\displaystyle #1$}\kern8pt}
 \kern8pt\vrule}\hrule}}
 %
 %
 

 \font\cmss=cmss10
 \font\cmsss=cmss10 at 7pt
 \def\rlx{\relax\leavevmode}
 \def\inbar{\vrule height1.5ex width.4pt depth0pt}
 \def\IC{\relax\,\hbox{$\inbar\kern-.3em{\rm C}$}}
 \def\IN{\relax{\rm I\kern-.18em N}}
 \def\IP{\relax{\rm I\kern-.18em P}}

\def\ZZ{\rlx\leavevmode\ifmmode\mathchoice{\hbox{\cmss Z\kern-.4em Z}}
  {\hbox{\cmss Z\kern-.4em Z}}{\lower.9pt\hbox{\cmsss Z\kern-.36em Z}}
  {\lower1.2pt\hbox{\cmsss Z\kern-.36em Z}}\else{\cmss Z\kern-.4em Z}\fi}
 \def\IZ{\relax\ifmmode\mathchoice
 {\hbox{\cmss Z\kern-.4em Z}}{\hbox{\cmss Z\kern-.4em Z}}
 {\lower.9pt\hbox{\cmsss Z\kern-.4em Z}}
 {\lower1.2pt\hbox{\cmsss Z\kern-.4em Z}}\else{\cmss Z\kern-.4em Z}\fi}
 \def\IZ{\relax\ifmmode\mathchoice
 {\hbox{\cmss Z\kern-.4em Z}}{\hbox{\cmss Z\kern-.4em Z}}
 {\lower.9pt\hbox{\cmsss Z\kern-.4em Z}}
 {\lower1.2pt\hbox{\cmsss Z\kern-.4em Z}}\else{\cmss Z\kern-.4em Z}\fi}

 \def\narrowplus{\kern -.04truein + \kern -.03truein}
 \def\narrowminus{- \kern -.04truein}
 \def\narrowminussub{\kern -.02truein - \kern -.01truein}

 \def\frac#1#2{{#1\over #2}}

 \def\IZ{\relax\ifmmode\mathchoice
 {\hbox{\cmss Z\kern-.4em Z}}{\hbox{\cmss Z\kern-.4em Z}}
 {\lower.9pt\hbox{\cmsss Z\kern-.4em Z}}
 {\lower1.2pt\hbox{\cmsss Z\kern-.4em Z}}\else{\cmss Z\kern-.4em Z}\fi}
 \def\IB{\relax{\rm I\kern-.18em B}}
 \def\IC{{\relax\hbox{$\inbar\kern-.3em{\rm C}$}}}
 \def\Ic{{\relax\hbox{$\inbar\kern-.22em{\rm c}$}}}
 \def\ID{\relax{\rm I\kern-.18em D}}
 \def\IE{\relax{\rm I\kern-.18em E}}
 \def\IF{\relax{\rm I\kern-.18em F}}
 \def\IG{\relax\hbox{$\inbar\kern-.3em{\rm G}$}}
 \def\IGa{\relax\hbox{${\rm I}\kern-.18em\Gamma$}}
 \def\IH{\relax{\rm I\kern-.18em H}}
 \def\II{\relax{\rm I\kern-.18em I}}
 \def\IK{\relax{\rm I\kern-.18em K}}
 \def\IP{\relax{\rm I\kern-.18em P}}

 \font\cmss=cmss10 \font\cmsss=cmss10 at 7pt
 \def\IR{\relax{\rm I\kern-.18em R}}

 %

 %
 %
 \def\eqnn#1{\xdef
#1{(\secsym\the\meqno)}\writedef{#1\leftbracket#1}%
 \global\advance\meqno by1\wrlabeL#1}
 \def\eqna#1{\xdef
#1##1{\hbox{$(\secsym\the\meqno##1)$}}

\writedef{#1\numbersign1\leftbracket#1{\numbersign1}}%
 \global\advance\meqno by1\wrlabeL{#1$\{\}$}}
 \def\eqn#1#2{\xdef
#1{(\secsym\the\meqno)}\writedef{#1\leftbracket#1}%
 \global\advance\meqno by1$$#2\eqno#1\eqlabeL#1$$}

\lref\AK{
M.~Aganagic, A.~Karch, D.~Lust and A.~Miemiec,
``Mirror symmetries for brane configurations and branes at singularities,''
Nucl.\ Phys.\ B {\bf 569}, 277 (2000)
[arXiv:hep-th/9903093].}
\lref\morep{J. Edelstein, K. Oh, and
R. Tatar, ``Orientifold, Geometric Transition and Large N Duality for SO/Sp
Gauge Theories,'' JHEP {\bf 0105} (2001) 009  [arXiv:hep-th/0104037]\semi
K. Dasgupta, K. Oh, and R. Tatar, ``Geometric Transition, Large N Dualities
and MQCD Dynamics,'' [arXiv:hep-th/0105066] \semi
K. Dasgupta, K. Oh, and R. Tatar,
``Open/Closed String Dualities and
Seiberg Duality from Geometric Transitions in
    M-theory,'' [arXiv:hep-th/0106040].}
\lref\AV{M. Aganagic and C. Vafa, ``Mirror symmetry, D-branes and
counting holomorphic discs,'' hep-th/0012041.}
\lref\AKV{M. Aganagic,
A. Klemm and C. Vafa, ``Disk instantons, mirror symmetry and the duality
web,'' hep-th/0105045.}
\lref\vafaleung{N.~C.~Leung and C.~Vafa,
``Branes and toric geometry,''
Adv.\ Theor.\ Math.\ Phys.\  {\bf 2}, 91 (1998)
arXiv:hep-th/9711013.}
\lref\gopvi{R.~Gopakumar and C.~Vafa,
``M-theory and topological strings. II,''hep-th/9812127.}
\lref\zasgr{T.~Graber and E.~Zaslow,
``Open string Gomov-Witten invariants: Calculations and a mirror  'theorem',''
hep-th/0109075.}
\lref\HV{K.~Hori and C.~Vafa,``Mirror symmetry,''
arXiv:hep-th/0002222.}
\lref\ach{B.~S.~Acharya,
``On realising N = 1 super Yang-Mills in M theory,''arXiv:hep-th/0011089.}
\lref\tayv{T.~R.~Taylor and C.~Vafa,
``RR flux on Calabi-Yau and partial supersymmetry breaking,''
Phys.\ Lett.\ B {\bf 474}, 130 (2000), hep-th/9912152.}
\lref\syz{A.~Strominger, S.~T.~Yau and E.~Zaslow,
``Mirror symmetry is T-duality,''
Nucl.\ Phys.\ B {\bf 479}, 243 (1996),hep-th/9606040.}
\lref\awi{M.~Atiyah and E.~Witten,
``M-theory dynamics on a manifold of G(2) holonomy,''
arXiv:hep-th/0107177.}
\lref\phases{E.~Witten,``Phases of N = 2 theories in two dimensions,''
Nucl.\ Phys.\ B {\bf 403}, 159 (1993), [arXiv:hep-th/9301042].}
\lref\klebstra{
I.~R.~Klebanov and M.~J.~Strassler,
``Supergravity and a confining gauge theory: Duality cascades and  chiSB-resolution of naked singularities,''
JHEP {\bf 0008}, 052 (2000) [arXiv:hep-th/0007191].}
\lref\BCOV{
M.~Bershadsky, S.~Cecotti, H.~Ooguri and C.~Vafa,
``Kodaira-Spencer theory of gravity and exact results for quantum string amplitudes,''
Commun.\ Math.\ Phys.\  {\bf 165}, 311 (1994),[arXiv:hep-th/9309140].}
\lref\may{
P.~Mayr,``On supersymmetry breaking in string theory and its realization
in brane  worlds,'',Nucl.\ Phys.\ B {\bf 593}, 99 (2001)
[arXiv:hep-th/0003198].}
\lref\partc{
A.~Giveon, A.~Kehagias and H.~Partouche,
``Geometric Transitions, Brane Dynamics and Gauge Theories,''
arXiv:hep-th/0110115.}
\lref\partb{
P.~Kaste, A.~Kehagias and H.~Partouche,
``Phases of supersymmetric gauge theories from M-theory on G(2)  manifolds,''
JHEP {\bf 0105}, 058 (2001)
[arXiv:hep-th/0104124].
}
\lref\parta{H.~Partouche and B.~Pioline,
`Rolling among G(2) vacua,''
JHEP {\bf 0103}, 005 (2001)
[arXiv:hep-th/0011130].}
\lref\amer{
A.~Iqbal and A.~K.~Kashani-Poor,
``Discrete symmetries of the superpotential and calculation of disk  invariants,''arXiv:hep-th/0109214.}
\lref\hiv{K.~Hori, A.~Iqbal and C.~Vafa,``D-branes and mirror symmetry,''
arXiv:hep-th/0005247.}
\lref\gomis{
A.~Brandhuber, J.~Gomis, S.~S.~Gubser and S.~Gukov,
``Gauge theory at large N and new G(2) holonomy metrics,''
Nucl.\ Phys.\ B {\bf 611}, 179 (2001)
[arXiv:hep-th/0106034].
}
\lref\cvetic{
M.~Cvetic, G.~W.~Gibbons, H.~Lu and C.~N.~Pope,
``M3-branes, G(2) manifolds and pseudo-supersymmetry,''
arXiv:hep-th/0106026.}
\lref\kmcg{
S.~Kachru and J.~McGreevy,
``M-theory on manifolds of G(2) holonomy and type IIA orientifolds,''
JHEP {\bf 0106}, 027 (2001)
[arXiv:hep-th/0103223].}
\lref\hose{
J.~D.~Edelstein and C.~Nunez,
``D6 branes and M-theory geometrical transitions from gauged  supergravity,''
JHEP {\bf 0104}, 028 (2001)
[arXiv:hep-th/0103167].
}

\Title
 {\vbox{
 \baselineskip12pt
 \hbox{HUTP-01/A048}
 \hbox{hep-th/0110171}\hbox{}\hbox{}
}}
 {\vbox{
 \centerline{$G_2$ Manifolds, Mirror Symmetry and Geometric Engineering}
 \vglue .5cm
 \centerline{}
 \vglue .5cm
 \centerline{}
 }}
 \centerline{ Mina ${\rm Aganagic}$ and
Cumrun ${\rm Vafa}$}
 \bigskip\centerline{ Jefferson Physical Laboratory}
 \centerline{Harvard University}
\centerline{Cambridge, MA 02138, USA}
 \smallskip
 \vskip .3in \centerline{\bf Abstract}
{We construct Calabi-Yau geometries with wrapped D6 branes
which realize ${\cal N}=1$ supersymmetric $A_r$ quiver theories,
and study the corresponding geometric transitions.
This also yields new
large $N$ dualities for topological strings
generalizing topological strings/large $N$ Chern-Simons duality.
  Lifting up to M-theory yields
smooth quantum geometric transitions without branes or fluxes, in the
context of $G_2$ holonomy manifolds.
In addition we construct a linear sigma model realization which is relevant
for the worldsheet theory of
superstrings propagating in local manifolds with $G_2$ holonomy, and
obtain mirror geometries
 for this class of supersymmetric sigma models.}
 \smallskip
\Date{October 2001}
\newsec{Introduction}
Type II strings compactified on Calabi-Yau threefolds
have been an extremely fruitful arena for interaction between
string theory and quantum field theory.  In the context of ${\cal N}=1$
supersymmetric theories this involves in addition wrapped spacetime filling
branes or dual descriptions involving fluxes through non-trivial cycles.

Both type IIA and type IIB pictures have proven to be useful in this regard:
Type IIB is useful because the dual descriptions involving fluxes yield
exact results for the vacuum geometry of the corresponding field theory
as there are no relevant instantons to modify the classical picture.
This idea has been shown to be rather powerful in
a number of examples
\lref\vaug{C.~Vafa,
``Superstrings and topological strings at large N,''
hep-th/0008142.}
\lref\civ{F.~Cachazo, K.~A.~Intriligator and C.~Vafa,
``A large N duality via a geometric transition,''
Nucl.\ Phys.\ B {\bf 603}, 3 (2001), hep-th/0103067.}
\lref\cfet{F.~Cachazo, B.~Fiol, K.~A.~Intriligator, S.~Katz and C.~Vafa,
``A geometric unification of dualities,''hep-th/0110028.}
\lref\ckv{F.~Cachazo, S.~Katz and C.~Vafa,
``Geometric transitions and N = 1 quiver theories,''hep-th/0108120.}
\refs{\vaug,\klebstra,\civ,\ckv,\cfet, \morep }.
On the other hand the type IIA description has also proven
to be useful.  The Seiberg-like dualities are often more manifest
in that formulation \lref\oogv{H.~Ooguri and C.~Vafa,
``Knot invariants and topological strings,''
Nucl.\ Phys.\ B {\bf 577}, 419 (2000),hep-th/9912123.}
\lref\ovseib{H.~Ooguri and C.~Vafa,
``Geometry of N = 1 dualities in four dimensions,''
Nucl.\ Phys.\ B {\bf 500}, 62 (1997), arXiv:hep-th/9702180.}
\refs{\ovseib, \cfet}\ as there are no worldsheet
instantons modifying the structure of Lagrangian cycles that D6 branes wrap.
Moreover the lift of this picture to M-theory gives the large N dualities/
gaugino condensation phenomena
a completely geometric description without branes or fluxes
\lref\amv{M.~Atiyah, J.~Maldacena and C.~Vafa,
``An M-theory flop as a large N duality,''
hep-th/0011256.}
\lref\AVG{M.~Aganagic and C.~Vafa,``Mirror symmetry and a G(2) flop,''
arXiv:hep-th/0105225.}
\refs{\amv,\ach,\AVG,\awi}.
One goal of this paper is to provide a type IIA setup
which mirrors the type IIB constructions of \refs{\civ,\ckv,\cfet}.  Along
the way we will propose some large $N$ dualities which are relevant
for open topological strings, generalizing the one proposed in
\ref\gopv{R.~Gopakumar and C.~Vafa,
``On the gauge theory/geometry correspondence,''
Adv.\ Theor.\ Math.\ Phys.\  {\bf 3}, 1415 (1999),hep-th/9811131.}
to some other non-trivial examples.  Furthermore, we
find a linear sigma model realization of superstring propagation
in the background of local $G_2$ holonomy manifolds. Moreover we
derive a mirror description of the same geometry.

Geometric engineering of similar ${\cal N}=1$ QFT's in the
context of M-theory on $G_2$ holonomy manifolds has recently been
considered in \partc .

\newsec{Review of geometric transition with branes}

In \vaug , by embedding
the large $N$ Chern-Simons/topological string duality \gopv\ in superstring
theory, the theory of $N$ D6 branes theory on the
$S^3$ in ${\cal M}_C = T^*S^3$ was conjectured to be dual
to theory on the small resolution of the conifold
$M_{K}=O(-1)+ O(-1) \rightarrow \bf{P}^1$
with $N$ units of RR flux through the $\bf{P^1}$ .
Let $s$ be the (complexified) size of the $\bf{P^1}$ and $Y={V\over g_s}+iC$
where $V$ is the size of
the $S^3$ the D6 branes wrap and $C$ the corresponding
vev of the 3-form gauge field on $S^3$.
 It was shown in \vaug\ that $s$ and $Y$ are related by
$$1-e^{-s}=e^{-Y/N}.$$
At large $N\gg 1$, the ${\bf P}^1$ grows a finite size.
The large $N$ duality is a geometric transition, as the
two geometries ${\cal M}_C$ and ${\cal M}_K$ are related by
conifold transition.

Below we will consider generalization of this duality to more
complicated geometries in type IIA compactifications. We will
subsequently make contact with geometric transitions and gauge
theory dualities in the context of type IIB string theory
compactifications studied in \refs{\civ,\ckv,\cfet}. We begin by
reviewing the geometry of \vaug\ .

\subsec{A geometric transition}
We can describe ${\cal M}_K$ as a Higgs branch of a $U(1)$ gauge
theory with four supercharges, and four chiral fields $x_i, i=1,\ldots,4$
\phases\ .
The D-term potential is minimized at
$$|x_1|^2+|x_4|^2-|x_2|^2-|x_3|^2=s$$
so that $x_{1,4}$ have charge $+1$ and $x_{2,3}$ charge $-1$, under the
$U(1)$. Here and in the following, the FI parameters $s$, when
combined with the corresponding $\theta$-angle become complexified.
We continue to call the complex quantity by the same symbol. Of course
the real part of that quantity appears in the D-term equation
(i.e. it is understood in the above formula that we mean $Re(s)$ rather
than $s$ on the right hand side).
Now consider the complex structure of ${\cal M}_K$.
Let $x,y,u,v$ be the gauge invariant variables,
$x=x_1x_2,\;y=x_3x_4,\;u=x_1x_3$ and $v=x_2x_4$.
They satisfy one relation
\eqn\consing{xy=uv.}
We can view ${\cal M}_K$ as a ${\bf C^*}\times{\bf C^*}$ fibration over the
$z$ plane
$$xy=z,\quad uv =z,$$
where fibers degenerate at $z=0$.
The two $U(1)$ isometries of ${\bf C^*}\times{\bf C^*}$,
correspond to
$$x,y,u,v\rightarrow xe^{i\theta_b},ye^{-i\theta_b},ue^{i\theta_a},
ve^{-i\theta_a},$$
and act on the $x_i$'s as:
$(x_1,x_2,x_3,x_4) \rightarrow (x_1 e^{-i (\theta_a-\theta_b)}, x_2 e^{i
\theta_a},x_3e^{-i\theta_b},x_4)$.
This resolves the fixed point set as follows:
\bigskip
\centerline{\epsfxsize 3.0truein\epsfbox{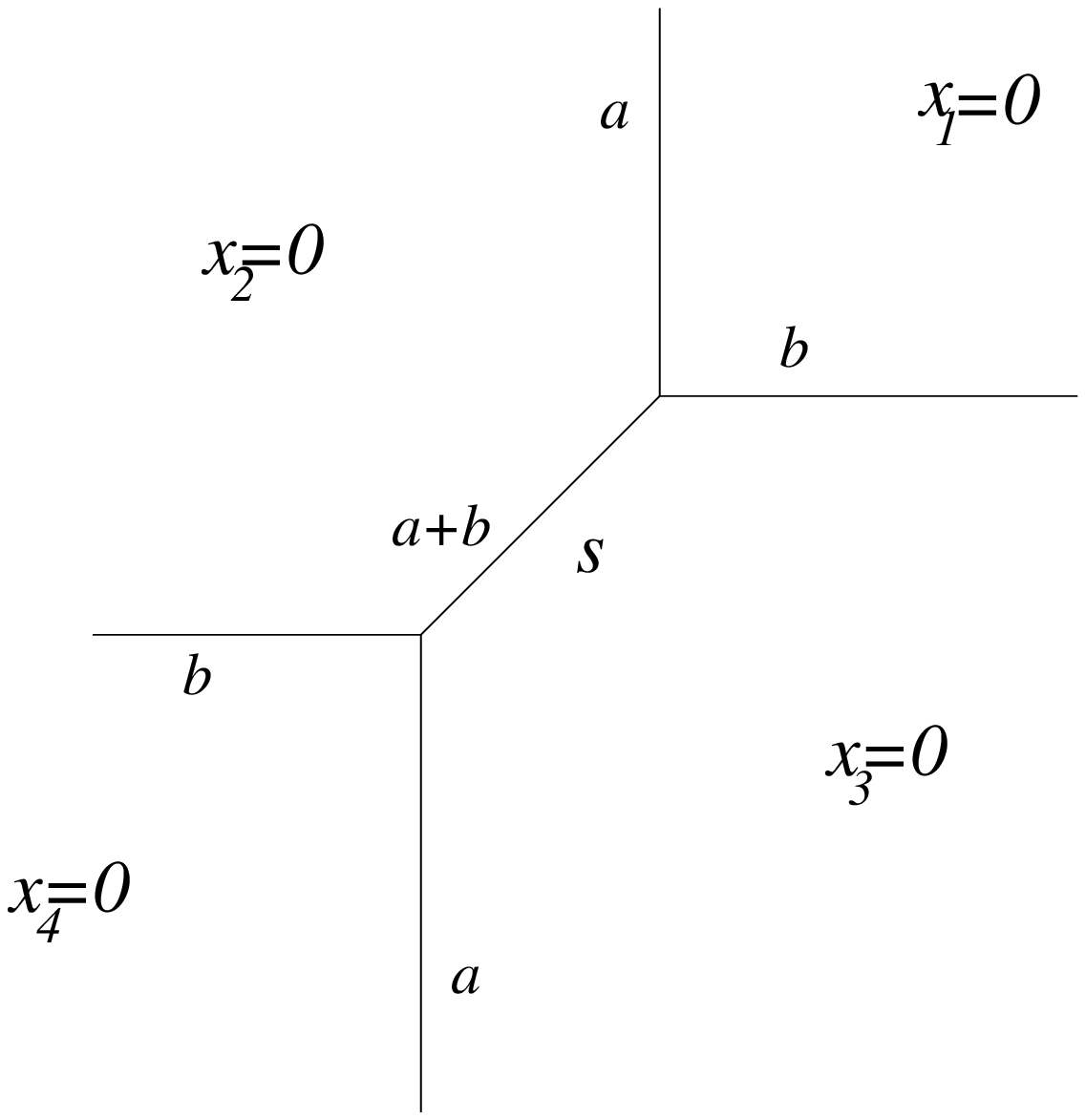}}
\rightskip 2pc \noindent{\ninepoint\sl
\baselineskip=8pt {\bf
Fig.1}{  The figure depicts the fixed point locus of the $U(1)$
isometries. Both degenerate over $z=x_1 x_2 x_3 x_4=0$.
For example, $U(1)_a$ degenerates over $x=0$ at
$y$ arbitrary, which is a line
$|x_4|^2=|x_3|^2+s$, and
$y=0$ and $x$ arbitrary where $|x_1|^2=|x_2|^2+s$. }}
\bigskip
We can alternatively consider ${\cal M}_K$ as a $T^3$ fibration
over ${\bf R^3}_{+}$, as in \refs{\vafaleung,\AV}.
The base of the fibration
is a space of solutions to D-term equation with $|x_i|^2$
as variables, out of which three are independent.
The fiber is a torus of phases of $x_i$ modulo the gauge group.
The $T^2\subset {\bf C^*}\times {\bf C^*}$ together with another phase
which can be identified with that of
$z$ are the $T^3$ fiber in the description of \AV\ .
Putting $|x_i|^2=r+c_i$ for all $i$ solves the D-term equations for any $r$
if $c_i$'s do, and the radial coordinate $r$ corresponds to $|z|$ \AK .
We will subsequently use the two pictures interchangeably .

Taking  $s$ to zero, the ${\bf P^1}$ in ${\cal M}_K$ goes to zero size
and the manifold is singular.
Continuation to negative values of $s$ leaves us within the moduli space
of ${\cal M}_K$ as the physics is controlled by the complexified
$s$, and so the singularity can be avoided
\lref\greenemor{P.S. Aspinwall, B.R. Greene
and D.R. Morrison,``Multiple mirror manifolds and topology change
in string theory,'' Phys. Lett. {\bf B303} 249 (1993), [arXiv:hep-th/9301043].}
\refs{\phases,\greenemor}. While the original
${\bf P}^1$ has negative area, there is a different curve whose volume is
positive. This is a flop.

\bigskip
\centerline{\epsfxsize 3.5truein\epsfbox{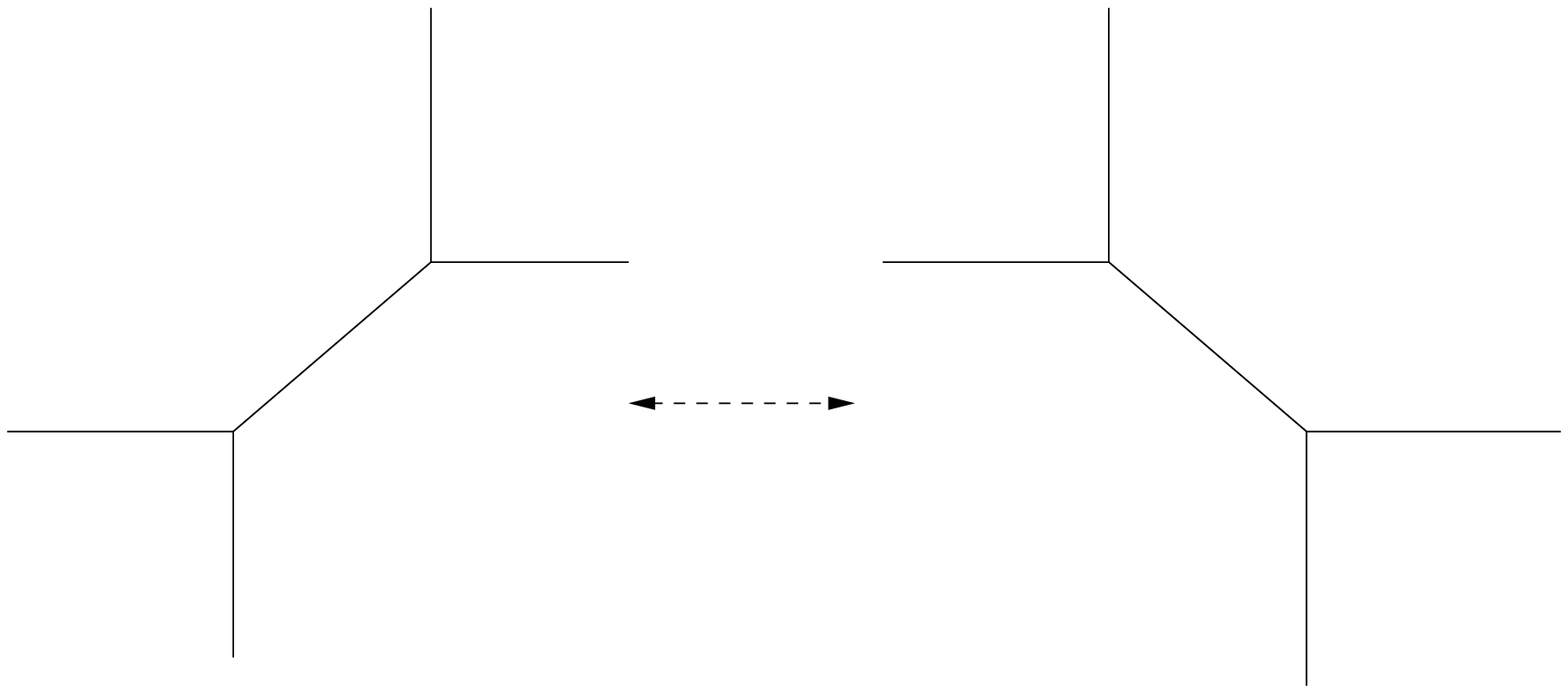}}
\rightskip 2pc \noindent{\ninepoint\sl
\baselineskip=8pt {\bf
Fig.2a}:{ This figure depicts a toric base of a flop where
one ${\bf P}^1$ shrinks and another one grows.}}
\bigskip

However, at $s=0$, a new phase opens up via a transition to a topologically
distinct manifold ${\cal M}_C=T^*S^3$, obtained by deforming \consing\ to
$$xy=uv +a$$
The fixed point set of $U(1)_a$ and $U(1)_b$
do not intersect any more:

\bigskip
\centerline{\epsfxsize 3.0truein\epsfbox{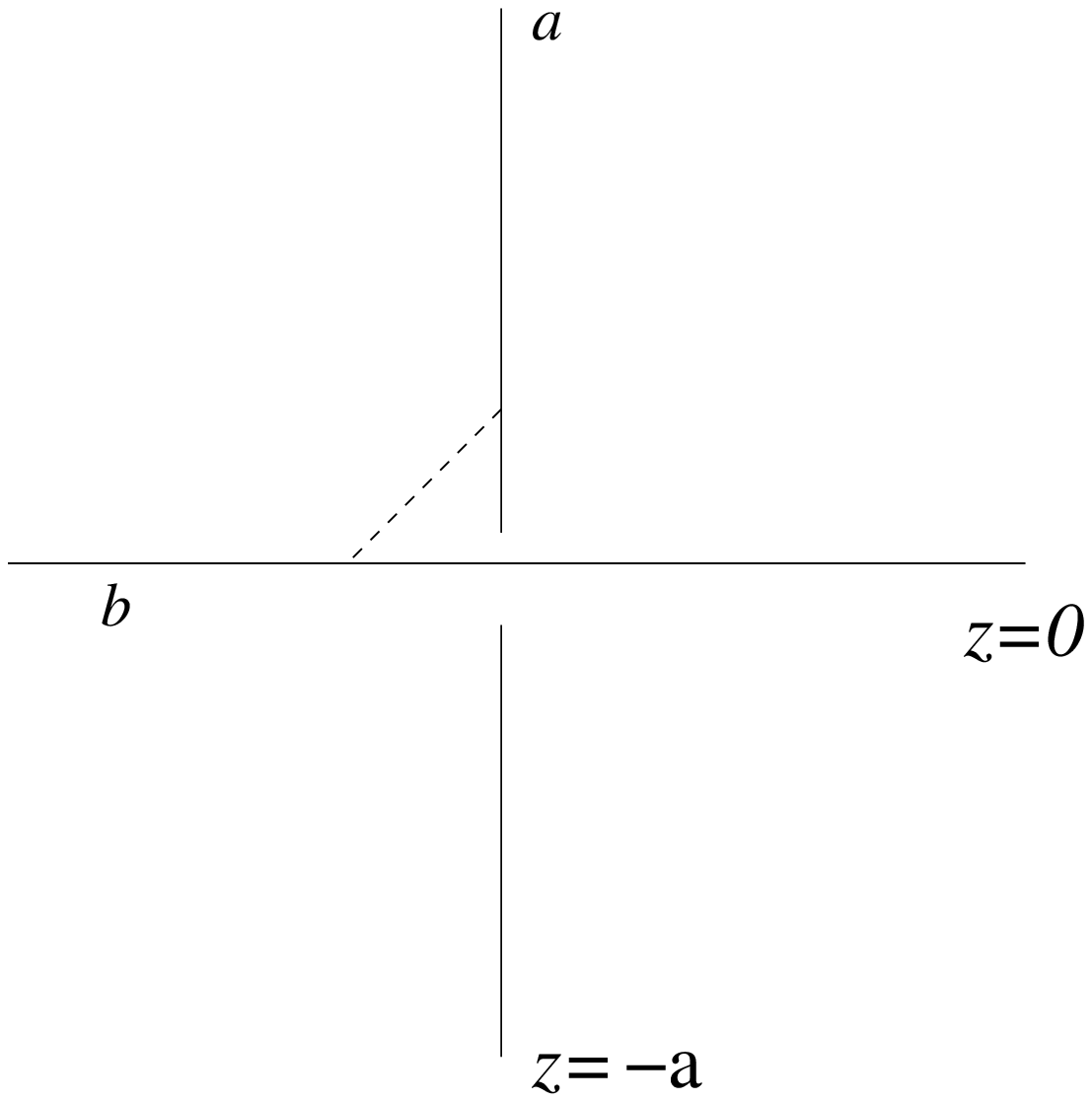}}
\rightskip 2pc \noindent{\ninepoint\sl
\baselineskip=8pt {\bf
Fig.2b}:{Writing $xy=uv+a$ as $xy=z+a\;, uv=z$,
the fixed point locus of $U(1)_a$, $x=0=y$ is at $z=-a$.
Also, $U(1)_b$ degenerates over $u=0=v$ which is at $z=0$. }}
\bigskip

Putting $y=\bar x, v=-\bar u$, we see that ${\cal M}_C$ contains an $S^3$
given by $|x|^2+|u|^2=a$. The $S^3$ is a line in the $z$
plane between $z=0$ and $z=-a$, together with the
$\theta_a$ and $\theta_b$ circles that degenerate over its endpoints.

\subsec{Large $N$ duality}

Let $S$ denote the expectation value of the gaugino bilinear of
the ${\cal N}=1$ SYM theory on the D6 brane,
$S=\langle Tr {{\cal W}}^2 \rangle$,
where ${\cal W}$ is the gaugino superfield. At low energies, the
$SU(N)$ subgroup of the $U(N)$ theory confines, and there is a gaugino
condensate, $S\neq 0$.
As shown in \refs{\BCOV,\vaug}\ , this generates
an effective superpotential of the four dimensional theory given by
\eqn\efsa{W^{\cal O}(S) = N \sum_{h=1}^\infty h {F}_{0,h}^{\cal O}S^{h-1},}
where $F^{\cal O}_{0,h}$ is the
topological $A$ model
open string partition function at genus-zero with $h$-holes.

On the other hand \refs{\tayv,\may}\
the RR two-form flux through the $S^2$
of the small resolution generates a superpotential given by
\eqn\efsb{W^{\cal C}(s)  = N\frac{\partial {F}_{0}^{\cal C}}{\partial s}}
where $F^{\cal C}_{0}$ is the
genus zero partition function of the closed string topological $A$ model.
Here $s$ is the complexified volume of the $\bf{P^1}$.
It was shown in \vaug\ that the duality of \gopv\ can be
interpreted by identifying $S=s$ which would yield
\eqn\ln{W^{\cal O} = W^{\cal C}.}
This relation has been further elaborated in \AVG .

\newsec{Transition $(h_{1,1},h_{2,1})=(2,0)\rightarrow (1,1)$}
Let us now consider a more complicated situation by embedding
the above transition in a geometry that is more complex.
Consider  ${\cal M}_K$, realized by linear sigma model,  given as:
\eqn\egone{\eqalign{|x_1|^2&+|x_4|^2 - |x_2|^2-|x_5|^2 =s,\cr
                    |x_3|^2&+|x_5|^2-2|x_4|^2 = t,}}
modulo the $U(1)^2$ actions, whose charges can be read off from the
above equations as being given by
$$Q_s = (1,-1,0,1,-1),\quad Q_t = (0,0,1,-2,1).$$
There are two Fayet-Iliopolous parameters $s,t$ controlling the
geometry, so $h_{1,1}({\cal M}_{K}) =2$.
The geometry of the ${\bf C}^*\times \bf{C}^*$ fibration is as follows:

\bigskip
\centerline{\epsfxsize 3.5truein\epsfbox{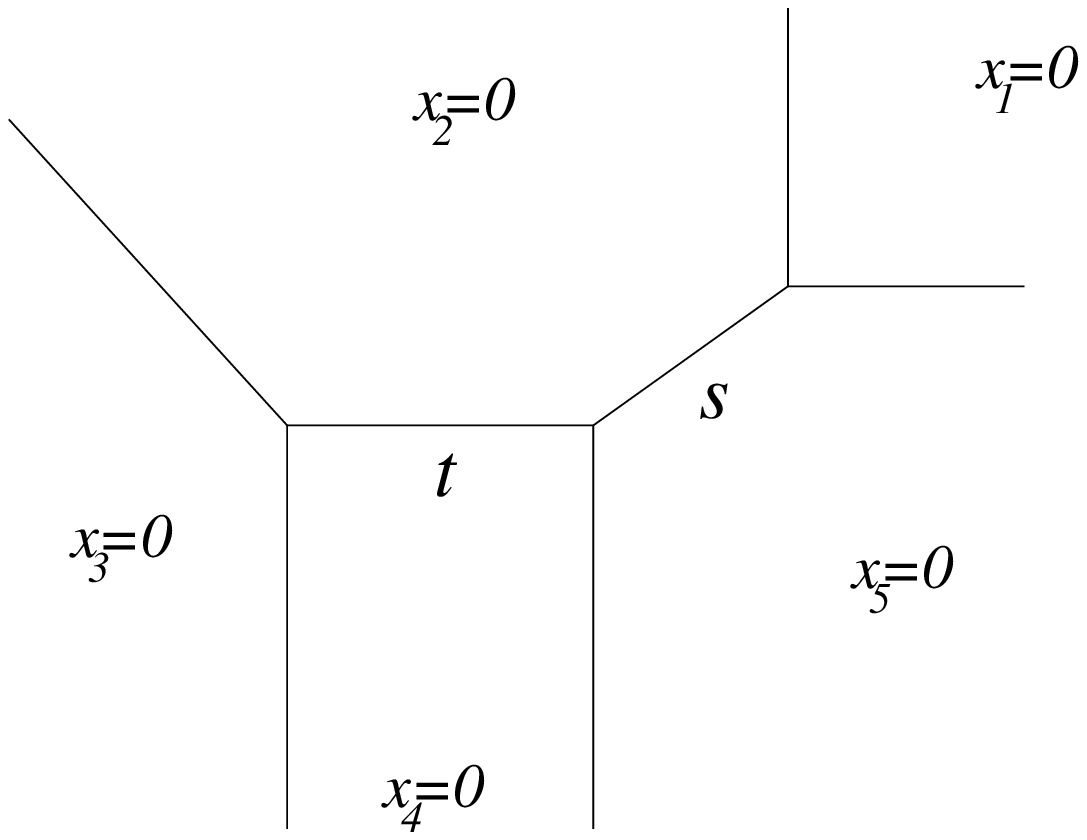}}
\rightskip 2pc \noindent{\ninepoint\sl
\baselineskip=8pt {\bf
Fig.3}: The toric projection of the geometry of a local
3-fold with $(h_{1,1},h_{2,1})=(2,0)$. }
\bigskip

We can reduce the geometry here
to the one we discussed in the previous
section in the $t\rightarrow \infty$ limit.
The parameter $s$ controls the volume ${\bf P^1}$ whose normal bundle is
$O({-1})+ O(-1)$ geometry.
${\cal M}_K$ is smooth as long as $s$ and $t$ are not zero.
For $s=0$, ${\cal M}_K$ develops a conifold singularity at
$x_1=0=x_2=x_4=x_5$.
The complex structure of ${\cal M}_K$ is
$$uv = xy^2,$$
where $x,y,u,v$ are gauge invariant coordinates,
$$x = x_1 x_2,\;\; y = x_3 x_4 x_5, \;\; u= x_1 x_4  x_5^2,\;\; v=x_2 x_3^2
x_4.$$
Let $\rho=u/y=x_1x_5/x_3$, which is gauge invariant and
good variable for $x_3\neq0$. In terms of $\rho$ we can write ${\cal M}_K$
as
$$v\rho = xy,$$
so in this patch the conifold
singularity is visible. The transition to ${\cal M}_C$
is made by deforming the complex structure of the local singularity as
\eqn\df{v\rho  = xy+a.}
The deformed manifold ${\cal M}_C$  contains a single $S^3$ whose size
is set by $|a|$. The manifold has $(h_{1,1},h_{2,1})=(1,1)$.

\bigskip
\centerline{\epsfxsize 4.0truein\epsfbox{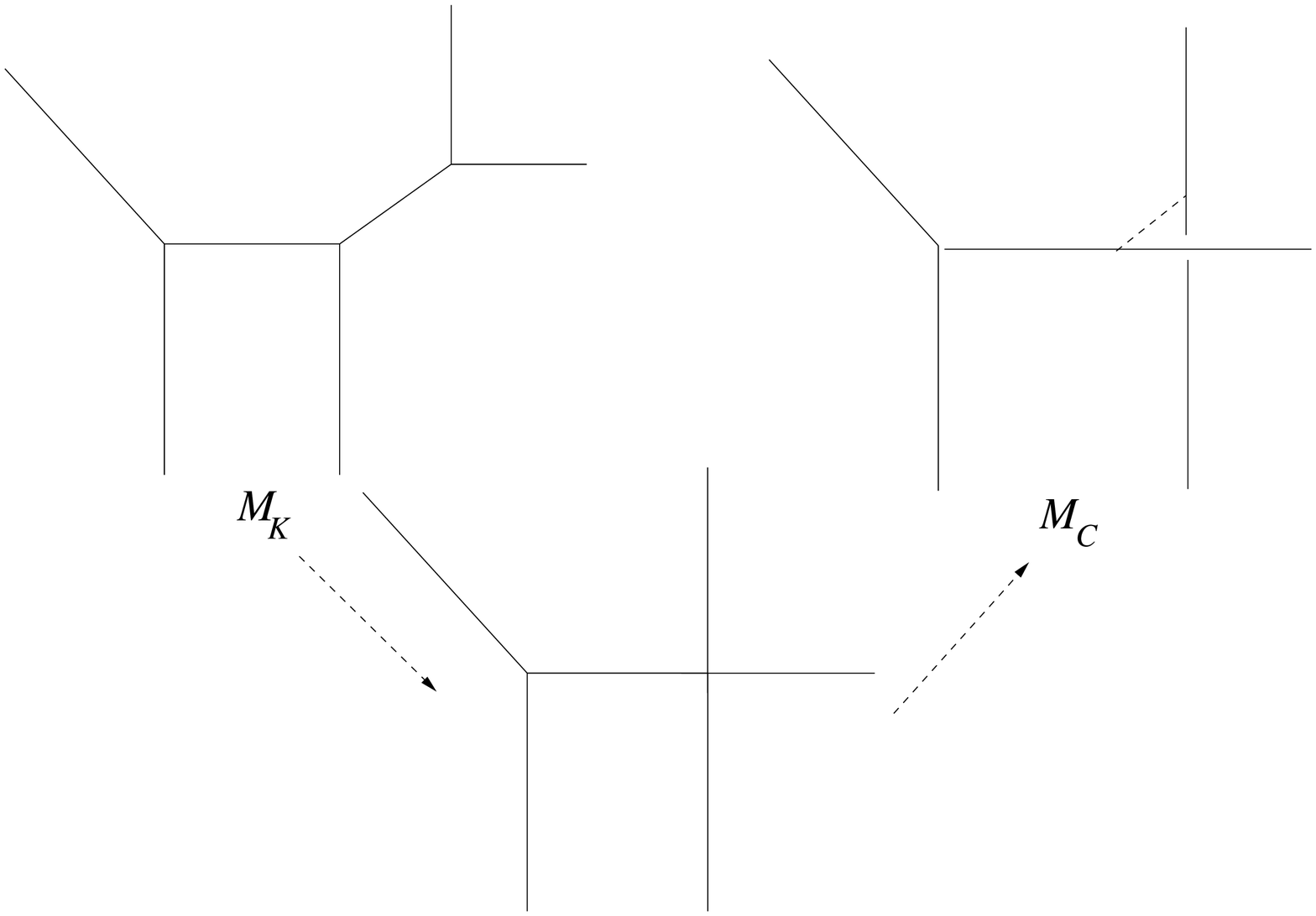}}
\rightskip 2pc \noindent{\ninepoint\sl
\baselineskip=8pt {\bf
Fig.4}: Blowing down the manifold ${\cal M}_K$ with $(h_{1,1},h_{2,1})=
(2,0)$ to a singular manifold and deforming the manifold to
${\cal M}_C$ with $(h_{1,1},h_{2,1})=(1,1)$ .}
\bigskip

\subsec{D6 branes on $S^3$ in ${\cal M}_C$}

We consider $N$ D6 branes on $S^3$ in ${\cal M}_C$.
The $S^3$ is isolated so
the massless fields on the D6 brane are described by a four dimensional
${\cal N}=1$ SYM theory with gauge group $U(N)$.
However, while the geometry is locally that of $T^*S^3$,
the global geometry of the manifold is more complicated. In particular,
$h^{1,1}({\cal M}_C)\neq 0$, and this can affect the A-model amplitudes.

The A-model amplitudes are given in terms of holomorphic maps of
string world-sheets into the Calabi -Yau, and with boundaries on the
D-branes. In the case of $T^*S^3$ the only such maps are the degenerate ones,
coming from the boundaries of the moduli space.

This may appear to be the answer for any finite value of $t$ as
well. This is because while ${{\cal M}_C}$ has a Kahler modulus
parameterized by $t$, there is no holomorphic curve representing
this class. We can write $\rho$ in \df\ as $\rho=\rho_1/\rho_2$
with homogeneous coordinates $(\rho_1,\rho_2 )\in {\bf P^1}$, so
\eqn\blow{\eqalign{&v\rho_1 =(xy+a)\rho_2,\cr &u\rho_2=y\rho_1}.}
It follows that a curve which is holomorphic near its north pole,
$\rho \sim 0$, blows up near the south pole, and vice versa. For
example, one can identically satisfy the last of the two equations
in \blow , everywhere on the curve, by putting $u=y=0=x$, but then
we have $$v = a/\rho,$$ which has a pole in the at $\rho=0$ if $a$
is not zero.

However, while the holomorphic $\bf{P^1}$ does not exist when the $S^3$ is
finite size,
in the presence of $D^6$ branes on the $S^3$, there are holomorphic $disks$
in ${\cal M}_C$ with boundary on the $S^3$.
The $S^3$ is given by $v=\bar\rho$ and $x=-\bar y$ in ${\cal M}_C$.
There is a disk at $u=0=y=x$, given by $|\rho|^2\geq a$ that
has a boundary at $|\rho|^2=a$ on the $S^3$ which is holomorphic.
By deformation of this disk to a closed sphere, while
ending on the Lagrangian
D6 brane (as can be seen from Fig. 5) it follows that it  has  Kahler volume
given by $t$.

\subsec{Relation to D6 branes in ${\bf C}^3$}

Note that in the limit where the size $|a|$
of the $S^3$ is taken to infinity, the Lagrangian
submanifold degenerates to ${\bf{C}}
\times S^1$ and ${\cal{M}_C}$ to ${\bf C^3}$.
In this limit one can expect to reproduce the results of \AV\ ,
where disk instantons were found to
generate superpotential.

To explain this consider the embedding of the
$S^3$ in ${\cal M}_C$ in more detail (see Fig.5)
Two of the $U(1)$ isometries of ${\cal M}_C$,
are generated by $(x,y,\rho,v) \rightarrow (x,y, \rho e^{-i \theta_a}, ve^{i
\theta_a})$ and $(x,y,\rho ,v) \rightarrow (xe^{i
\theta_b}, ye^{-i \theta_b}, \rho ,v)$.
For  $z = xy=\rho v -a$, the $\theta_a$ and $\theta_b$ circles
degenerate over $z=-a$ and
$z=0$ respectively,  and together with a line in $z$ plane connecting the two
points form the $S^3$.

Thus, near $z=xy \sim 0$, the $S^3$ looks like $\bf{C\times S^1}$,
where the $S^1$ corresponds to $\theta_a$. Taking the fixed point
of this isometry, located at $z=-a$, to infinity, the
configuration is exactly that of \refs{\AV,\AKV}. The primitive
holomorphic disk that generated the superpotential in \AKV\ goes
over to the holomorphic disk constructed above for the finite size
$S^3$ (see Fig. 5). Furthermore, since the two configurations
differ by variation of complex structure, the parameter $a$, the
$A$ model amplitudes in the two cases should be related, modulo
potential contributions from boundaries of moduli spaces
(degenerate maps which ``feel'' the infinity on $S^3$).

In this geometry there is in fact a whole family of Lagrangian 3-cycles, with
topology of  $S^3$ or $S^2\times S^1$ which is depicted in Fig. 5.

\bigskip
\centerline{\epsfxsize 4.0truein\epsfbox{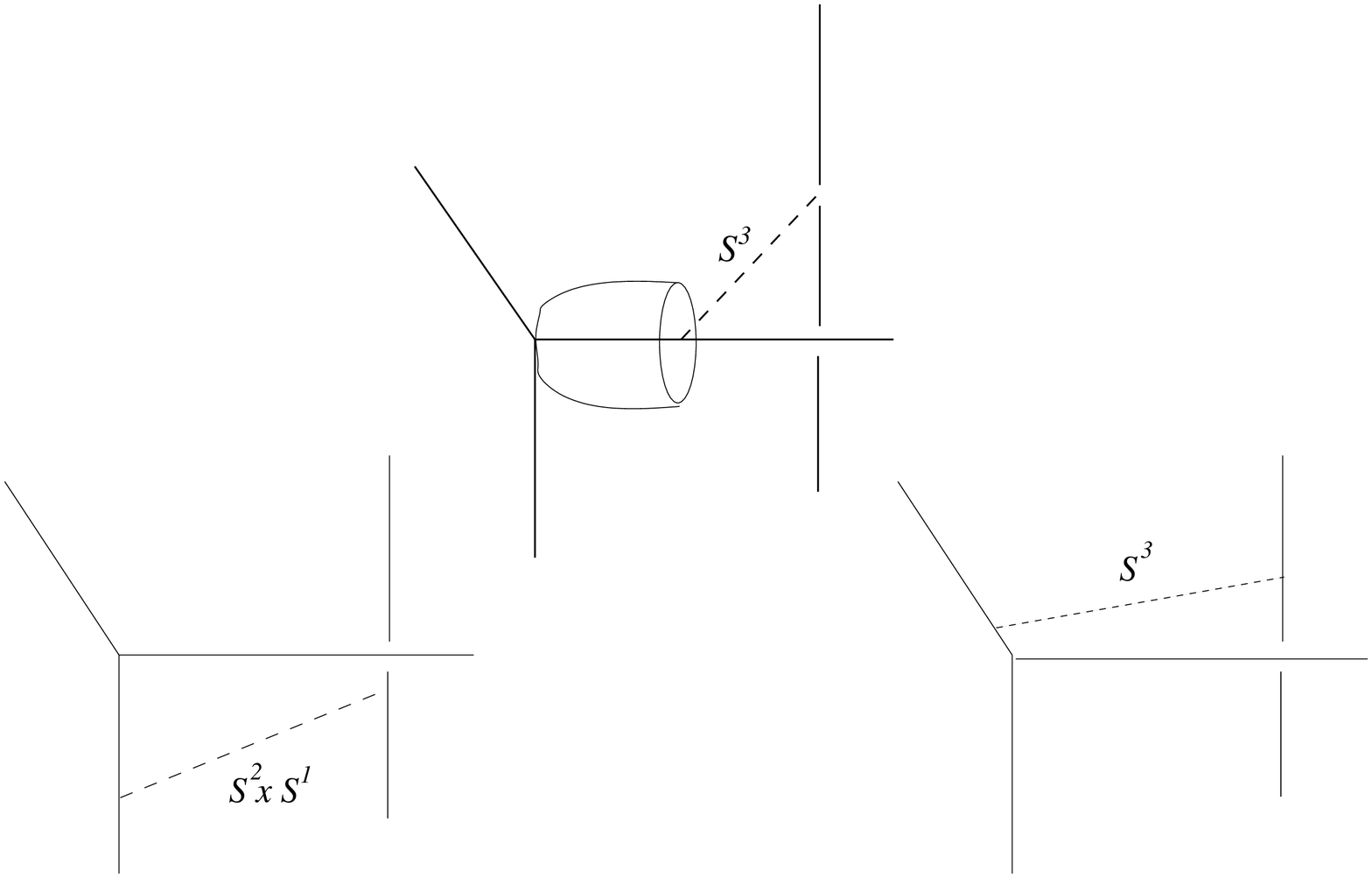}}
\rightskip 2pc \noindent{\ninepoint\sl
\baselineskip=8pt {\bf
Fig.5}:There is a family of  Lagrangian 3 cycles with topology
$S^3$ and $S^2\times S^1$ that can be deformed to each other.
In the quantum theory (taking worldsheet instantons into
account) they are smoothly connected. }
\bigskip

The Lagrangian submanifolds can be deformed to each other continuously
and in particular they represent the same 3-cycle class, by varying
the height on the ladder it ends on (see Fig. 5).
Note that classically the configurations look very different, and do
not represent {\it special} Lagrangian submanifolds.  Roughly
speaking minimal volume of the brane
projects to the shortest length
on the toric base.   The only one which is classically stable is
the
$S^3$ depicted at the top of Fig. 5.    The class
 with topology $S^2\times S^1$ is almost stable, namely
 if we consider sliding the brane far down the ladder it
 becomes supersymmetric (as if it is wrapping an $S^2\times S^1$ in $K3
 \times {\bf C^*}$).
Note that the one with the topology of $S^3$ has no moduli, whereas
the Lagrangian submanifold with topology of $S^2\times S^1$ has a
one-dimensional moduli space (as expected from the fact that $b_1=1$
\syz ).  This one dimensional moduli
space of $S^2\times S^1$ corresponds to sliding it up and down
the ladder in the above figure.  As we will discuss
in the next section, when we consider the effects of worldsheet instantons,
 all the above
branes are smoothly connected as we vary moduli. Moreover
for any given moduli, there is a unique stable brane, which
in the classical limit gets identified with the $S^3$ drawn
at the top of Fig. 5.

\newsec{D5 branes on the mirror ${\cal W}_C$}

In \refs{\AV ,\AKV }\
we studied mirror symmetry of D6 branes wrapped on non-compact special
Lagrangian submanifolds of topology ${\bf C}\times S^1$ on ${\bf C^{3}}$.
We have seen in the previous section that this theory arises as a limit
of D6 brane theory on $S^3$ and the one on $S^2\times S^1$
in ${\cal M}_C$ in a suitable limit.
 Mirror symmetry allowed us to compute the disk
amplitude for the D6 branes using topological $B-$model, where the
superpotential is computed from classical geometry. In this
section we will use mirror symmetry to compute the $A$ model disk
amplitude for the present case by borrowing the techniques in
\refs{\AV, \AKV}.

The mirror of ${\cal M}_C$ can be derived using the known mirror of
${\cal M}_K$, and following the mirror pair through the conifold transition.
The mirror manifold of ${\cal M}_K$ is given
by \HV\
$${\cal W}_K:\quad xz= 1+e^{-u} + e^{-v} + e^{u-t}+ e^{-v + u-t-s}.$$
Here $t$ and $s$ are complex structure moduli mirror to the complexified
Kahler parameters of ${\cal M}_K$.
The physical, flat coordinates $\hat t,\hat s$ which measure
the BPS D-brane tensions are in this case given by
$e^{-t} = e^{-{\hat t}} f^{-2}$ and
$e^{-s} = e^{-{\hat s}}f$, for $f= 1+e^{-{\hat t}}$  . In terms of these,
the equation for ${\cal W}_K$ is given by
\lref\peter{P. Mayr,``N=1 Mirror Symmetry and Open/Closed String Duality,''
[arXiv:hep-th/0108229].}
\lref\indian{S. Govindarajan, T. Jayaraman, T. Sarkar,``Disc Instantons
in Linear Sigma Model,''[arXiv:hep-th/0108234].}
\eqn\wk{{{\cal W}_K}:\quad xz =(1-e^{-\hat u})(1-e^{\hat u-\hat t})+
e^{-\hat v}(1-e^{\hat u-\hat t-\hat s}).}
where $\hat u$ and $\hat v$ are the open string counterparts of
flat coordinates \refs{\AKV}
(see also the related discussions in
\refs{\amer,\peter,\indian}),  $e^{-u} = e^{-{\hat u}}f^{-1}$ and
$e^{-v}=e^{-\hat v}f^{-1}$.
In writing \wk\ we have absorbed a factor of $f$ in $z$.
This  has a conifold singularity at $\hat s = 0$ where the
manifold degenerates as
$$ xz =(1-e^{-\hat u}-e^{-\hat v})(1-e^{\hat u-\hat t}),$$
which is singular at $x=0=z=1-e^{-\hat u}-e^{-\hat v}$ and $\hat
u=\hat t$. We can blow this up to get a smooth manifold ${\cal
W}_C$, by replacing the conifold singularity with a $\bf{P^1}$,
parameterized by homogeneous coordinates $\rho_1,\rho_2$
\eqn\mblow{\eqalign{&x\rho_2=\rho_1( 1-e^{-\hat u}-e^{-\hat v})\cr
&z\rho_1=\rho_2(1-e^{\hat u-\hat t})}}
In terms of the inhomogeneous coordinates, this gives ${\cal W}_C$
as a hypersurface
$$x\lambda = 1-e^{-\hat u}-e^{-\hat v},$$
in a patch with coordinates $x,\lambda=\rho_2/\rho_1,\hat u,\hat v$.
In the other patch, with $y,\lambda'=1/\lambda,\hat u ,\hat v$ as coordinates,
we have
$$y \lambda'=1-e^{\hat u-\hat t},$$
and the transition functions are inherited from \mblow .
A D5 brane wrapping
the holomorphic ${\bf P^1}$ above is
mirror to the D6 brane on the $S^3$ in ${\cal M}_C$.

\subsec{ Disk Amplitude}
The superpotential can be computed as in \refs{\AV,\ckv}.
We have that $$W = \int_{B} \Omega,$$ where $\Omega$ is the holomorphic
threeform,
$$\Omega = \frac{dz}{z} du dv,$$
and $B$ is a three-chain which moves the holomorphic curve to a homologous
one,
see Fig.6. The homologous 2-cycle is mirror to the Lagrangian
brane with topology of $S^2\times S^1$, and very far down the ladder
it has a normal bundle which is ${\cal O}(-2)+{\cal O}(0)$ and becomes
supersymmetric.
  So in the very setup of this
computation we see that the mirror connects the two different kinds
of Lagrangian submanifolds smoothly.
The north and the south pole of the curve move on
the two Riemann surfaces
\eqn\poles{\hat u=\hat t,\;\;\; 0=1-e^{-\hat u}-e^{-\hat v}.}
%

\bigskip
\centerline{\epsfxsize 3.0truein\epsfbox{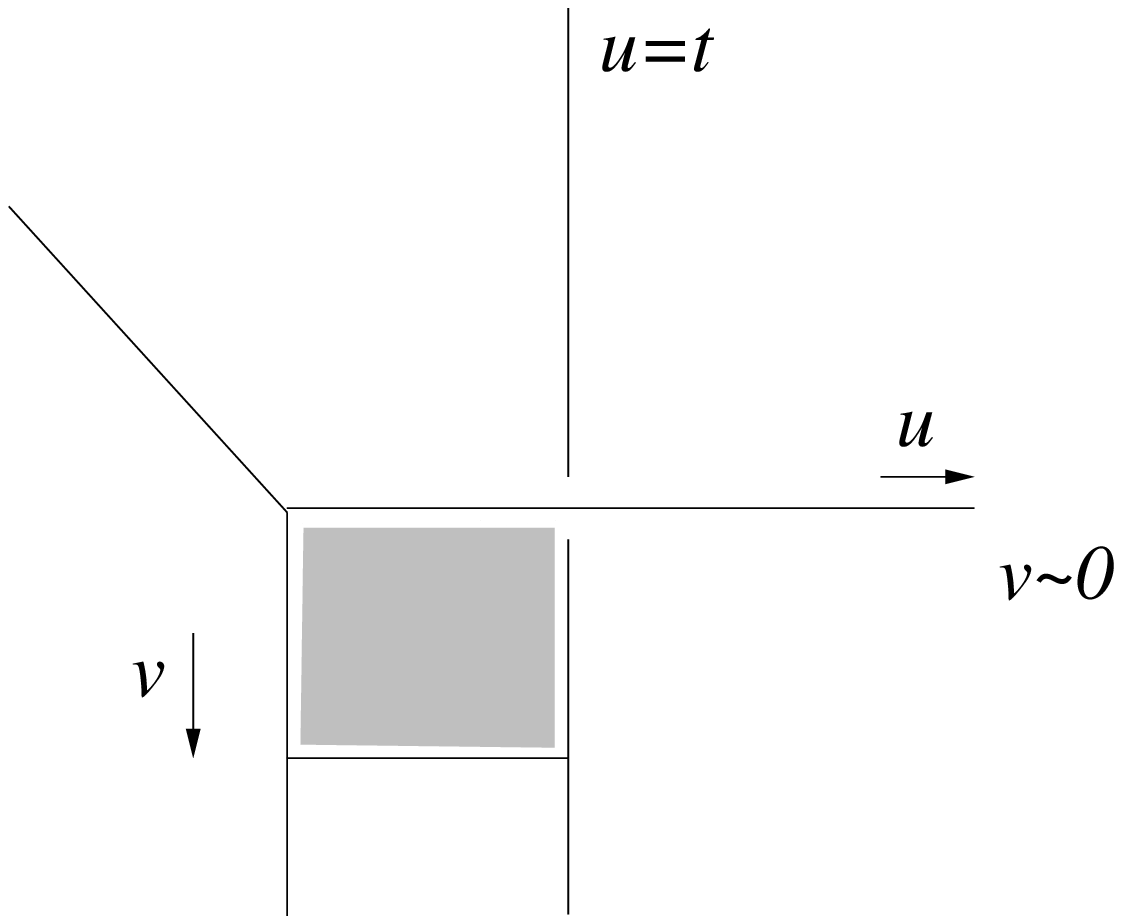}}
\rightskip 2pc \noindent{\ninepoint\sl
\baselineskip=8pt {\bf
Fig.6}: The computation of the superpotential $W$ reduces
to the integral of a 1-form on the boundary of
the shaded region $D$. }
\bigskip

In the following for simplicity of notation we drop the hats off
the $u,v,t$.
We can take the curves in the chain
to be parameterized by $v$ (see Fig. 6). The superpotential evaluates to
$$W =\int_{B} \frac{dz}{z} du dv = \int_{\partial D} u dv,$$
where $D$ is the shaded region in Fig.6. This gives
\eqn\suppo{W = \int log(1-e^{-v}) dv +t\;v.}
 The critical point of $W$ is given by
$$dW=0\rightarrow (1-e^{-v})=e^{-t}.$$
Note that in the classical limit, $t\gg 0$, the mass of $v$
is $\sim e^{t}$, so $v$ is fixed at the minimum of the superpotential.
The critical point of $W$ freezes $v$ at $v\sim 0$,
which is what one expects for the mirror of the $S^3$ Lagrangian brane.
In this context the existence
of a frozen $v$ is consistent with the fact that the $S^3$
Lagrangian submanifold is not expected to have any moduli.

On the other hand if we consider the regime $v\gg 0$, we have $W\sim t v$.
This is linear in $v$ and
can be viewed as an ${\cal N}=2$ FI term, which is consistent
with the fact that in this limit the brane is the mirror of the $S^2\times
S^1$ Lagrangian submanifold.  Furthermore, in the region with $v\ll 0$, we
have $W\sim -v^2/2 +tv$  which reflects the fact that there
are no classical special Lagrangian submanifolds there.  Note, however,
that in the non-classical regime (from the A-model viewpoint), where
$t\ll 0$ these branes become stable at  $v\sim t$.  Thus we
see that all the different classes of Lagrangian submanifolds are
smoothly connected in the quantum theory.

Note that for N D5 branes on the $S^2$,
$v$ is the scalar in the adjoint-valued chiral
multiplet and $tr W(v)$  is the tree-level superpotential of the $U(N)$
theory on the brane.

\subsec{Disk amplitude in the mirror A-model}

As discussed above, by mirror symmetry the equation \suppo\
gives a tree-level superpotential
of $N$ D6 branes on the $S^3$ in ${\cal M}_C$. This is a good
description in the classical domain where $t\gg 0$.
The minimum of the superpotential with respect to $v$, located at
\poles\ , coincides with the location of the holomorphic $\bf{P^1}$ in
\mblow\ . The value of the superpotential at the minimum is
\eqn\suppomin{W(v_{min}) = - \sum_{n=1} \frac{e^{-n t}}{n^2}.}
This is
found by noting that $$W(v_{min}) = \int^{v_{min}} u(v) dv - t\; v_{min} =
-\int^{t} v(u) du,$$ and integrating by parts.
As expected from the discussion of the previous section,
the formula for the disk amplitude
has the form which corresponds to a single primitive disk instanton
of size $t$, ending on the D6 brane.
\subsec{Geometric Transition and Large $N$ duality}

With $N$ D6 branes on $S^3$ in ${\cal M}_C$, the natural generalization of the
conjecture of \vaug\ is that the large $N$ dual is the theory
on ${\cal M}_K$ with Ramond-Ramond flux. The flux is turned on through
the $S^2$ in ${\cal M}_K$ that is related by the conifold transition to the
$S^3$ in ${\cal M}_C$ which the D6 branes wrap.
The D6 branes disappear, and their Ramond-Ramond flux is supported by
the $S^2$.

Topological A model open string amplitudes $F_{0,h}^{\cal O}$ on
${\cal M}_C$
can in general depend
on $t$, the complexified Kahler modulus of ${\cal M}_C$. In the previous
section we showed this explicitly for the disk partition function $F_{0,1}$.
On the other hand, ${\cal M}_{K}$ has two Kahler moduli
$s,t$, and closed topological $A$ model amplitudes depend on both.
This then implies that \ln\ naturally generalizes to
\eqn\lng{\sum_{h=1}^\infty
F^{\cal O}_{0,h}(t)S^{h} = F^{\cal C}_{0}(s,t),}
with $S=s$.
To check this statement we would need to compute $F^{\cal C}_0(s,t)$,
which we will now turn to.  On the left hand side, we have computed
$F^{\cal O}_{0,h}$ for $h=1$ and so we will be able to make a comparison
for that case.   Moreover we can
also compute, as we will discuss below $F^{\cal C}_g(s,t)$ for arbitrary
genus $g$, which yields a prediction for all the open topological string
amplitudes.

\subsec{The effective superpotential}

We will now compute the effective superpotential using the large $N$ dual
description in terms of ${\cal M}_K$ with $N$ units of the RR flux through
the $\bf{P^1}$ of size $s$. Using the large $N$ duality, this gives a
prediction for the open string amplitudes $F^{\cal O}_{0,h}$ for any
number of holes $h$ on ${\cal M}_C$, which we can check for the case
$h=1$ that we have computed above.

The superpotential on ${\cal M}_K$  generated by the Ramond-Ramond flux
is given by
$$W(s,t) = N\frac{\partial F_0(s,t)}{\partial s}$$
where here we denote the flat coordinates by $s,t$ (i.e., drop
the hatted notation).  The superpotential is
 $N$ times the (instanton corrected)
Kahler volume of the 4-cycle dual to the $\bf{P^1}$.
The way to compute this is to use mirror symmetry
and map it to a classical period of the holomorphic three-form on
${\cal W}_K$.
{}From the equation of ${\cal W}_K$, \wk\
and $\Omega = \frac{dz}{z}dudv$,
we compute
$$\partial_s F_0 = \int_{C_4} \frac {dz}{z} du dv
= \int_{t+s}^{\Lambda}v(u) du,$$ where $$v(u) =
log[\frac{1-e^{u-t-s}}{(1-e^{-u})(1-e^{u-t})}].$$ Above, $\Lambda$
is an infra-red cutoff, regulating the infinite volume of the
four-cycle.\ To the leading order in $\Lambda$, and up to a
numerical constant, this is
$$\partial_s F_0 = -s(\Lambda -  t - s)
- \sum_{n=1}^{\infty}\frac{e^{- n s}}{n^2}
- \sum_{n=1}^{\infty}\frac{e^{- n ( s+ t)}}{n^2}.$$

This structure for $F_0$ can be interpreted as predicting the
existence of two stable D2 branes one wrapped over the ${\bf P}^1$
with volume $s$ and the other as a bound state of branes wrapped
over two ${\bf P}^1$'s with volumes $s,t$ (see Fig. 3).  Note that
there is also another family of ${\bf P}^1$'s with volume given by
$t$, but the number of BPS states associated with that will be
determined only if one compactifies the model.  At any rate, the
derivative with respect to $s$, which is the only relevant
information needed for the superpotential is insensitive to it. In
fact, using the ideas in \gopvi\ , one can determine all the
higher genus contributions for this topological theory. This
yields
$${\cal F}^{\cal C}(t,s,g_{s})=\sum_{n>0}
\bigg[\frac{e^{- n s}}{n(2{\rm sin}(ng_s/2))^2}
+\frac{e^{- n (s+t)}}{n(2{\rm sin}(ng_s/2))^2}\bigg]$$
up to a potential addition of $s$ independent terms (and up to a polynomial
of degree 3 coming from genus 0 and 1).

\subsec{Large $N$ conjecture}

We now generalize the conjecture of \gopv\ to the
case at hand.  In fact the topological string computes more than
just the superpotential, and so it is natural to identify
all such quantities between the open and closed string side,
as in \vaug\ (where it was noted that the higher
genus topological amplitudes compute certain
F-type terms in the dual gauge theory).  In particular the statement is that the  topological
open string, which includes contributions from arbitrary holes
and genera, agrees with the corresponding genus contribution
of the closed topological string, namely
$${\cal F}^{\cal O}(t,N,g_s)=\sum_{g,h}{ F}^{\cal O}_{g,h} (t)N^h g_s^{2g-2+h}=
\sum_g {F}^{\cal C}_g g_s^{2g-2}(t,s)={\cal F}^{\cal C}(t,s,g_s)$$
where $s=N g_s$ as in the conjecture of \gopv .
However in this case,
unlike \gopv , we do not have a complete computation on both
sides to check the above predictions.

We can compare the superpotential computation we have done here
in the context of mirror symmetry, assuming the identification $s=S$
between the closed string and open string.  As for the open string computation
we have only explicitly computed the contribution from configurations with one
hole, which yields the superpotential \suppomin.  To compare contributions
from one hole (since $S$ counts the number of holes according to \lng )
we expand $ F_0$ to first order in $s$, which is equivalent
to putting $s=0$ for $\partial_s F_0$ and we obtain (up to
a constant addition)
$$\partial_s F_0(s=0,t)= - \sum_{n=1} \frac{e^{-n  t}}{n^2}=W(v_{min})$$
 which is thus a check of the above conjecture.

We can also ask about higher number of holes and genera.
On the open string side
these are not easy to compute with present techniques (though recent
progress in this direction may shed light on it \zasgr\ ).
At any rate using the above structure yields a prediction.
For example, if we consider genus 0,
we can now extract the prediction for ${ F}_{0,h}$,
from the coefficient of $s^{h-1}$ of $\partial F_0(s,t)$.
For general $h$ we find
$$F^{\cal O}_{0,h} = (-1)^h\sum_{n=1}^{\infty}\frac{n^{h-3}}{h!}
(1+e^{-n t}).$$
which would be interesting to verify.  The more general prediction
we are making is that
$${\cal F}^{\cal O}(N,t,g_s)={\cal F}^{\cal C}(s=Ng_s,t,g_s)$$
where
$${\cal F}^{\cal O}(N,t,g_s)=
\sum_{g,h} F^{\cal O}_{g,h}(t)g_s^{2g-2} (Ng_s)^h$$
$${\cal F}^{\cal C}(s=Ng_s,t,g_s)=\sum_{n>0}
\bigg[\frac{e^{- n s}}{n{(2 \rm sin}(ng_s/2))^2}
+\frac{e^{- n (s+t)}}{n(2{\rm sin}(ng_s/2))^2}\bigg]$$
(again up to an addition of a polynomial of degree 3).
  In particular $F_{g,h}(t)$
represents the topological string amplitude with genus $g$ and
with $h$ holes  ending on $S^3$.  It would
be interesting to verify this prediction.
 This includes the contribution
of the wrapping around the non-trivial class given by area $t$.  This
generalizes the conjecture of \gopv\ to the case at hand.

\newsec{Transition $(h_{1,1},h_{2,1})=(4,0)\rightarrow (2,1)$}

We now consider another local CY 3-fold, and
its associated transitions, which we will later use to
geometrically engineer ${\cal N}=2$ gauge theories deformed
to ${\cal N}=1$ by the addition of the cubic superpotential term
for the adjoint scalar.

Consider Calabi-Yau manifold ${\cal M}_K$ given by the D-terms
\eqn\egbb{\eqalign{&|x_0|^2+|x_1|^2 -|x_2|^2-|x_6|^2 = {s_{1}},
\quad |x_3|^2+|x_6|^2 -2|x_0|^2= {t_{1}},\cr
&|x_0|^2+|x_5|^2 -|x_4|^2-|x_6|^2 = {s_{2}},\quad
|x_2|^2+|x_4|^2 -2|x_0|^2= {t_{2}},}}
and modulo the corresponding $U(1)^4$ action given by the charges
for $(x_0,x_1,...,x_6)$
\eqn\chb{\eqalign{&Q_{s_1}=(1,1,-1,0,0,0,-1)\quad
Q_{t_1}=(-2,0,0,1,0,0,1)\cr
&Q_{s_2}=(1,0,0,0,-1,1,-1)\quad Q_{t_2}=(-2,0,1,0,1,0,0)}}
%
\bigskip
\centerline{\epsfxsize 3.0truein\epsfbox{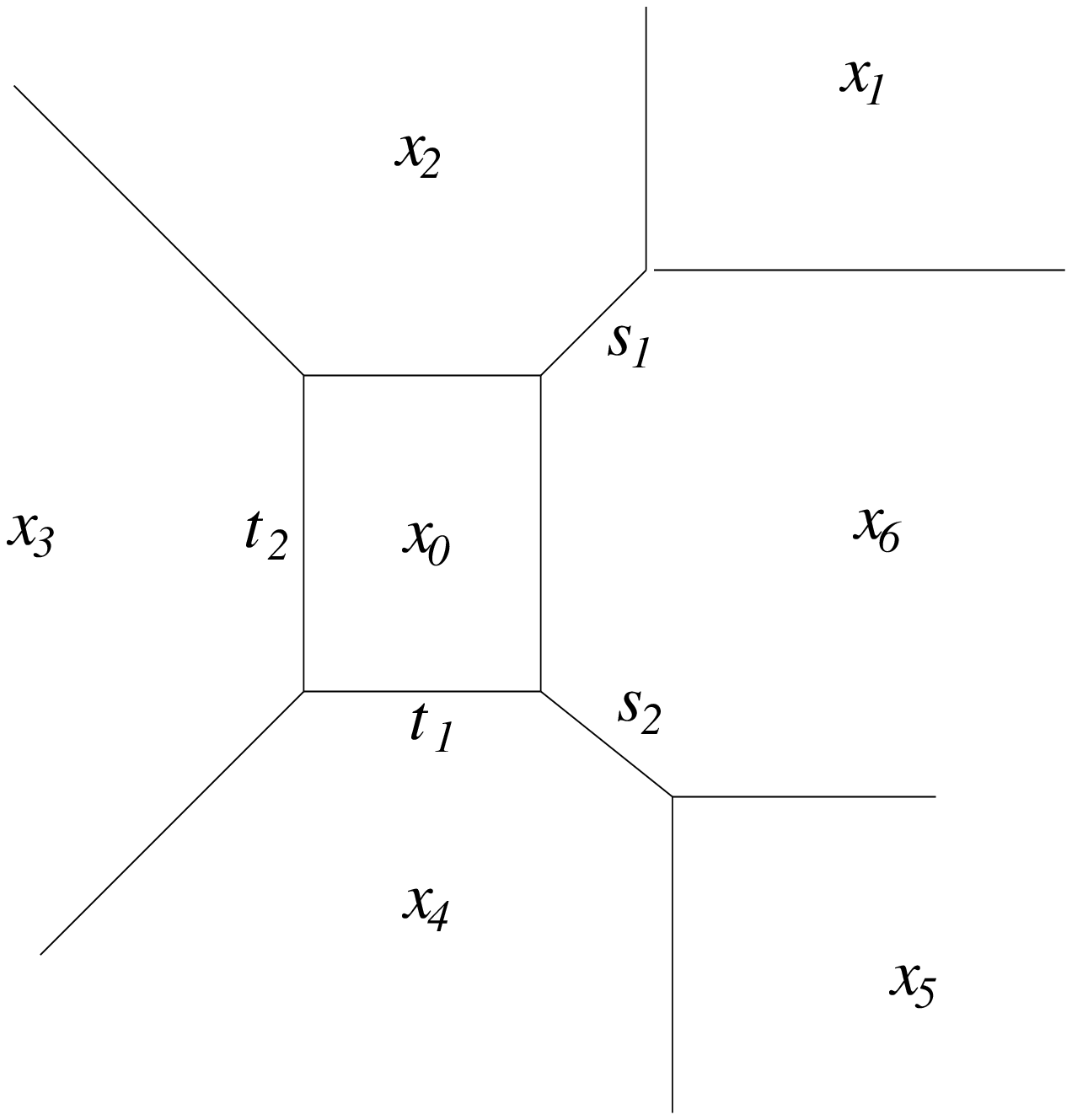}}
\rightskip 2pc \noindent{\ninepoint\sl
\baselineskip=8pt {\bf
Fig.7}: The toric projection of the geometry
of a local 3-fold with $(h_{1,1},h_{2,1})=(4,0)$.}
\bigskip
This has two conifold singularities as ${s_{1,2}}$ go to zero. In
fact, for ${t_2}$ very large, the local geometry in the
$x_{0,1,2,3,6}$ and $x_{0,3,4,5,6}$ coordinates is the one we
studied in the previous sections. One can therefore expect, that
one can continue beyond the singularity to a topologically
distinct manifold ${\cal M}_C$, with the two $S^2$'s replaced by
two $S^3$'s. This is true, apart from the fact that there is an
obstruction to blowing up the two $S^3$'s independently. This is
because in ${\cal M}_K$ there is a four-cycle $C$ represented for
example by $x_0=0$ locus. This four cycle intersects both $S^2$'s:
$$\#(C \cap S^2_1) = 1 = \#(C \cap S^2_2).$$
Upon contracting the $S^2_{1}$ and $S^2_{2}$ to zero size the four
cycle gets punctured by the two conifold singularities, and
blowing up the $S^3$'s it becomes a four-chain. Consequently, the
$S^3$'s are homologous and there is only one three-cycle class in
${\cal M}_C$. The transition is shown in figure 8.
\bigskip
\centerline{\epsfxsize 3.5truein\epsfbox{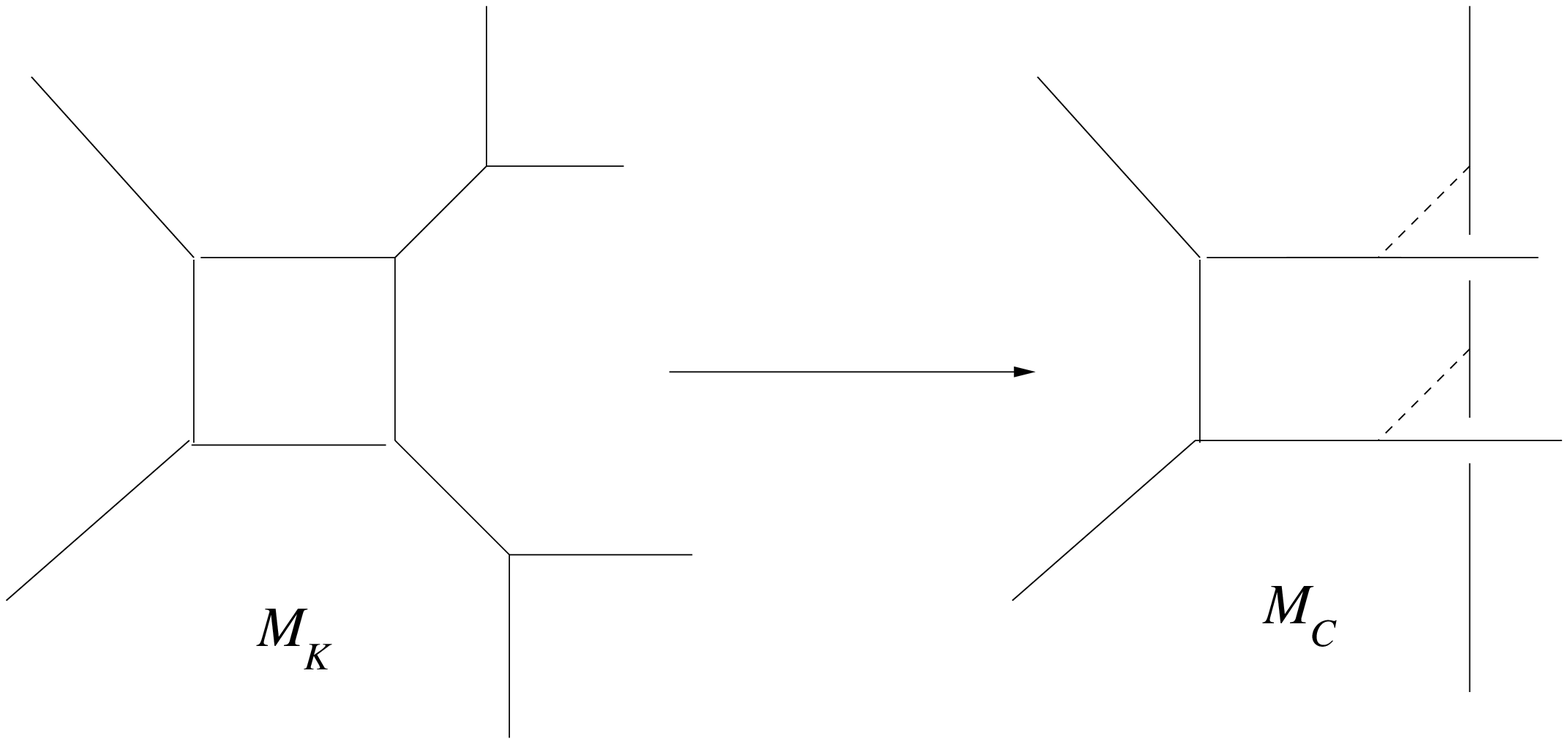}}
\rightskip 2pc \noindent{\ninepoint\sl
\baselineskip=8pt {\bf
Fig.8}: The transition of $(h_{1,1},h_{2,1})=(4,0)\rightarrow (2,1)$.
Note that the dashed lines correspond to two $S^3$'s in the same
$H_3$ class. }
\bigskip
\subsec{D branes and mirror symmetry} We can consider wrapping
some number $N_{1,2}$ D6 branes on the supersymmetric three-cycles
$S^3_{1,2}$. The massless fields on the D6 branes clearly give a
$U(N_1)\times U(N_2)$ gauge theory with ${\cal N}=1$ supersymmetry
and with no massless matter, for generic values of ${t_{1,2}}$.
The D6 brane theory, however, has at sufficiently high energies a
$U(N_1+N_2)$ gauge invariance, since the cost in energy in
bringing the two groups of D6 branes together is finite.
Furthermore, there is a massive adjoint chiral multiplet $v$
parameterizing the Lagrangian deformations of the three-cycles,
with a superpotential $Tr W(v)$ that has two vacua. In this
subsection, we will use mirror symmetry to compute the details of
$W(v)$.

The mirror manifold to ${\cal M}_K$ is given by
\eqn\wkb{{\cal W}_K:\quad xz = 1-e^{-u}-e^{-v} - e^{u-t_1} -
e^{v-t_2} + e^{-v+u-t_1-s_1}+ e^{v+u-t_2-t_1-s_2}.}
This has a singularity as $s_{1,2}$ approach zero (Note that
$s_{1,2}$ are not flat coordinates here. Consequently, we may only expect
$s_{1,2} \rightarrow 0$ for $t_{1,2}\rightarrow \infty$.)
where it degenerates as
\eqn\singb{xz = ((1-e^{-\hat v})(1-e^{\hat v-\hat t_2})-e^{-\hat
u})(1-e^{\hat u-\hat t_1}).}
Here, $e^{-t_1} = e^{-\hat t_1}(1+e^{-\hat t_2})/f^2$,
$e^{-t_2} = e^{-\hat t_2}/f^2$,
and $s_1,s_2$ are fixed in terms of these:
$$e^{-s_1} = f/(1+e^{-\hat t_2}) = e^{-s_2}.$$
We have defined $f=1+e^{-\hat t_1}+e^{-\hat t_2}$ and absorbed
a factor of $f$ in the definition of $z$.
In addition, we redefined $e^{-u} = e^{-\hat u} /f$,
$e^{-v} = e^{-\hat v}/f$, where $\hat u, \hat v$ are open string flat
coordinates in this locus in the complex structure moduli space.
Again for simplicity of notation we drop the hats off of the variables.
As before this geometry can be blown up to a smooth manifold ${\cal W}_C$
that is mirror to ${\cal M}_C$:
\eqn\blowb{\eqalign{x\rho_2& = \rho_1((1-e^{- v})(1-e^{ v-
t_2})-e^{- u})\cr z\rho_1&= \rho_2(1-e^{u-t_1}),}}
where $\rho_{1,2}$ are homogeneous coordinates on $\bf{P^1}$.

We can compute the tree-level superpotential $W(v)$
from the holomorphic Chern-Simons action as before:
\eqn\tsuppb{W(v) = \int u(v) dv =
- \int log[(1-e^{-v})(1-e^{v-t_2})]dv - t_1\;v.}
This corresponds in the D6 brane language to two primitive disk instantons
of actions $v$ and $t_2-v$ that wrap two ``halves'' of the
holomorphic $\bf{P^1}$
in ${\cal M}_C$, see figure 9.
\bigskip
\centerline{\epsfxsize 3.0truein\epsfbox{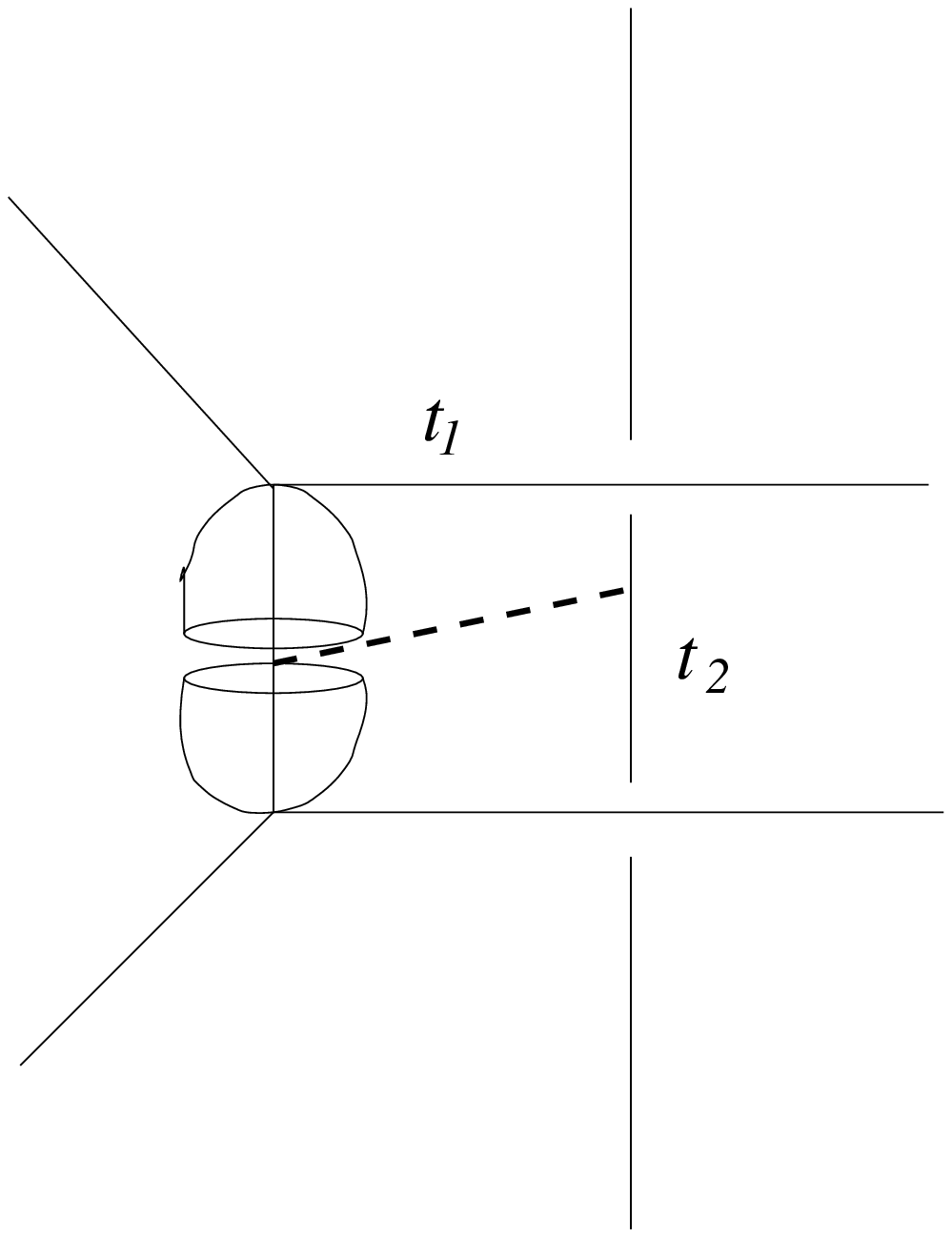}}
\rightskip 2pc \noindent{\ninepoint\sl
\baselineskip=8pt {\bf
Fig.9}{ The tree level superpotential for the adjoint scalar $v$
is generated by two primitive disk instantons, whose actions
are $v, t_2-v$. Integrating out $v$, this gives rise to
superpotential for $t_2$, generated by a single disk of size
$t_2$.}}
\bigskip
This has two minima at two solutions of
$$u=t_1, \quad e^{-u} = (1-e^{-v})(1-e^{v-t_2}).$$
Putting $N_{1,2}$ D5 branes on the two holomorphic ${\bf P^1}$'s
located at the two minima, we get the mirror of the $D6$ brane
configuration at its vacuum. Note that for large $t_1\gg 0$ the
minima are at $v=0$ and $v=t_2$ which are the classical locations
for the two mirror $S^3$'s.  For other values of $t_1$ the story
is similar to what we discussed in the context of the previous
example where the classical topology of the mirror brane changes.

\subsec{Large $N$ dual} Just as in the previous example one
naturally proposes a large $N$ duality where we wrap $N_1, N_2$ D6
branes around the two $S^3$'s in the same class and on the dual
side the branes disappear and are replaced by two $S^2$'s in
different classes of sizes $s_1,s_2$. The sizes of the two $S^2$'s
are fixed in terms of the two integers $N_1,N_2$ and the volume of
$S^3$, given by $Y$. The relation is found by extremizing the
superpotential
$$W_{eff} = \sum_{i=1,2} N_i \frac{\partial F_0}{\partial s_i} -Y (s_1+s_2).$$
The first term above is generated by $N_i$ units of
Ramond-Ramond flux through $S^2_i$ in ${\cal M}_K$
and $\frac{\partial F_0}{\partial s_i}$ is the volume of the dual
four-cycle.
Furthermore, $Y$ is the holomorphic volume of the dual three-cycle class
the D6 branes were wrapping, just as in \vaug .

This is the formulation of the large $N$ duality in the IIA superstring
setup.  However we can also recast it as a purely topological string duality.
In that context we have
$${\cal F}^{\cal O}(N_1,N_2,t_1,t_2, g_s)={\cal F}^{\cal
C}(s_1,s_2,t_1,t_2,g_s)$$
where $s_i=N_i g_s$ and ${\cal F}^{\cal O}$ and ${\cal F}^{\cal C}$ denote
the topological string amplitudes before (where we have D-branes)
and after transition (where branes disappear) respectively.  With the
present technology we can compute ${\cal F}^{\cal C}$ at least
at the level of genus 0 and with localization formula to higher genus as well,
though all genus computation is not available.  Similar comments apply
to the left hand side (using the techniques discussed in \zasgr ).  It would
be interesting to check the predictions made here.

\subsec{Geometric engineering of ${\cal N}=1$ QFT's}

We have seen in the above example that there are two minima for
the superpotential, which in the classical limit $t_1\gg0$ are
mirror to the two $S^3$'s, which are in the same class.
\lref\kkaa{S. Kachru, S. Katz, A. Lawrence, J. McGreevy, ``Open
string instantons and superpotentials,'' Phys. Rev. {\bf D62}
026001 (2000), [arXiv:hep-th/9912151].} This is a mirror situation
to the model studied in \refs{\kkaa,\civ }\ where there were two
$S^2$'s in the same class which were obtained by deformations
of a geometry
where the two $S^2$'s coalesced.  This was used to geometrically
engineer ${\cal N}=1$, $U(N)$ gauge theory with an adjoint field
$\Phi$, with  a cubic superpotential $W(\Phi)$. The two critical
points of $W$ corresponded to two holomorphic $S^2$'s and one can
distribute $N=N_1+N_2$ branes by wrapping $N_1$ around one $S^2$
and $N_2$ around the other.  This was found to have the
interpretation of the Higgsing $U(N)\rightarrow U(N_1)\times
U(N_2)$.

Here we wish to consider a limit where the above theory is realized
in the type IIA setup.  For this reason we have to consider the limit
where both $S^3$'s are very close to each other, or more precisely
in the quantum theory, where the critical points of the superpotential
are very close.  We will now consider such a limit to geometrically
engineer the above gauge theory in the context of type IIA.
Let us rewrite the superpotential \tsuppb\ in the form
$$\partial_{\tilde v} W={\rm log}{(1-\alpha e^{-{\tilde v}})(
1-\alpha e^{{\tilde v}})\over
\beta}$$
where
$${\tilde v}=v-{t_2\over 2}, \quad
\alpha =e^{-t_2/2}, \quad \beta =e^{-t_1}$$
It is easy to see that the condition that the two critical points
be the same is that
$$\beta =(1-\alpha)^2.$$
at which point both minima are at  ${\tilde v}=
v-{t_2\over 2}=0$.

We thus take
$$\beta =(1-\alpha)^2+\epsilon ^2$$
where $\epsilon \ll 1$.  Expanding the superpotential in this region,
near ${\tilde v}\sim O(\epsilon )$ we obtain
\eqn\cusu{{\partial}_{\tilde v} W={-1\over (1-\alpha)^2}[\alpha {\tilde v}^2+\epsilon
^2].}
In other words $W(\tilde v)$ is a cubic superpotential. Note that there
are two parameters controlling a cubic superpotential (up to shifting the
variable and addition of constant to $W$) and these are captured by
$\alpha$ and $\epsilon $ above.

In this limit the mirror geometry  ${\cal W}_C$
becomes
\eqn\local{xz = \tilde u((1-\alpha)^2 \tilde u -
\alpha \tilde v^2 - \epsilon^2).}
We used the fact that $u=t_1$ at the critical point to expand $u =
t_1 + \tilde u$ where $\tilde u \sim O(\epsilon^2)$. This
describes a local $A_1$ singularity in $\tilde u-x-z$ space
fibered over the $\tilde v$ plane. As a check, we can compute the
superpotential directly from \local\ . $W = \int \frac{dz}{z} du
dv = {-1\over (1-\alpha)^2}\int (\alpha \tilde v^2 + \epsilon^2)
dv$ in agreement with \cusu\ and the results of \civ .

At large $N$, the theory of D6 branes on ${\cal M}_C$ is
better described by geometry of ${\cal M}_K$ with Ramond-Ramond fluxes
turned on, as we discussed above.  Now we wish to consider the limit
of the above large $N$ duality in the context of geometric
engineering of ${\cal N}=1$ QFT's we are presently considering.

The limit of relevance, just as in \vaug\ is where the corresponding
blow ups $s_1,s_2$ are small.  This implies that the dual large $N$ description,
where the branes have been replaced by fluxes is given by
\eqn\singb{\eqalign{xz &= ((1-e^{-\hat v})(1-e^{\hat v-\hat t_2})-e^{-\hat
u})(1-e^{\hat u-\hat t_1})\cr
& - \delta s_1 e^{-\hat v + \hat u -\hat
t_1-s_1^0}  - \delta s_2 e^{-\hat t_2 +\hat v + \hat u -\hat
t_1-s_2^0}}}
where $s_{i} =s_i^0+ \tilde s_i$, and
$s^0_{1,2} =f/(1+\alpha^2)$ is the value of the non-flat
coordinate at the conifold point, as before.
This
carries over in the local limit to
deformation of \local\ as
$$xz = \tilde u((1-\alpha)^2 \tilde u -
\alpha \tilde v^2 - \epsilon^2) -
\alpha( \tilde s_1 +\tilde s_2 -(\tilde
s_1 -\tilde s_2)\tilde v).$$
The flat coordinates which to the leading order agree with
$s_{1,2}$ are dual to gaugino condensates corresponding to the
low-energy gauge group $U(N_1)\times U(N_2)$.  Note that the
deformations of the geometry corresponding to blowing up the
$S^3$'s in the mirror ${\cal W}_C$ involves terms which have
coefficients $1,{\tilde v}$ as anticipated in \civ\ by
normalizability of the corresponding harmonic form. {}From this
point on, the analysis of the large $N$ duality is identical to
that done in \civ\ in the context of type IIB strings.  In fact,
we can interpret the results of \civ\ as providing a check for a
particular limit of the type IIA duality we have conjectured here.

\subsec{Geometric Engineering of ${\cal N}=1$ $A_r$ Quiver
Theories} In this section we briefly outline the relevant type IIA
geometry which would realize the large $N$ dualities for ${\cal
N}=1$ $A_r$ quiver theories, studied in \refs{\ckv,\cfet}. These
can be viewed as deformations of ${\cal N}=2$ $A_r$ quiver
theories with addition of a polynomial superpotential  of degree
$p+1$ in the adjoint fields.  An example of relevant type IIA
geometry before and after transitions would look like that shown
in figure 10. The new aspect that arises for $r>1$ is that there
is bifundamental matter from strings stretching between D6 branes
wrapping three-cycles in different homology classes. We give an
example in the figure below for the case of $A_2$ quiver theory:

\bigskip
\centerline{\epsfxsize 4.5truein\epsfbox{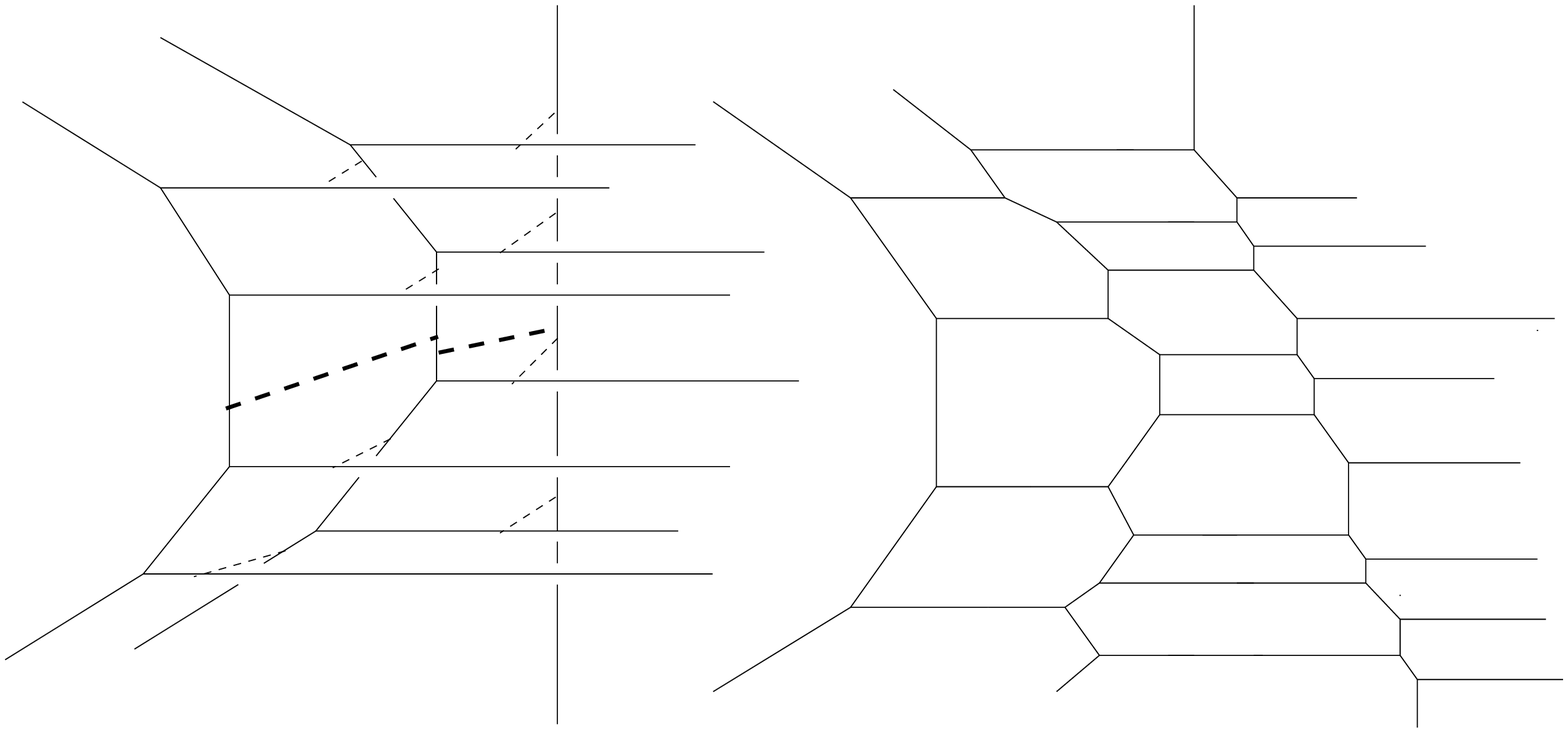}}
\rightskip 2pc \noindent{\ninepoint\sl \baselineskip=8pt {\bf
Fig.10}{  Type IIA geometry for $A_{r=2}$ quiver theory with a
superpotential that is quintic in the fields, for every node of
the quiver. The bold dashed lines represent Lagrangian $S^2\times
S^1$ cycles.  The four critical points of the superpotential for
each gauge factor corresponds to the four thin dashed lines.  Also
shown in the figure is the corresponding large $N$ dual with the
branes replaced by fluxes through the corresponding 2-cycles.}}
\bigskip
This has two classes in $H_3({\cal M}_K)$ and we can wrap $N_i$,
$i=1,2$ D6 branes on them. There are Lagrangian representatives of
this, that are of topology $S^2\times S^1$, and they engineer the
${\cal N}=2$ theory. There are $U(N_i)$ valued vector ${\cal N}=2$
vector multiplets, whose adjoint scalars move the three-cycles up
and down the ladder, just as in the $A_1$ case discussed above.
The hypermultiplets arise from three-cycles that can meet along an
$S^1$, and in this case there is a bifundamental in $(N_1,{\bar
N}_2)$. Supersymmetry is broken by superpotential generated by
disk instantons. Since there are $p=4$ vacua for each gauge group,
in appropriate local limit this engineers a quintic superpotential
for the adjoint chiral multiplet of each gauge group.

It is clear that we can repeat the same kind of analysis
we did above for the case of $A_1$ with a cubic superpotential.
We leave that to the interested reader.

\newsec{M-theory and geometric transitions}

Type IIA string theory compactified on a Calabi-Yau manifold ${\cal M}$
with Ramond-Ramond two form flux $[F]\in H^2({\cal M},{\bf Z})$ turned on
lifts to M-theory compactified on a
manifold of $G_2$ holonomy.
D6 branes wrapped on special Lagrangian submanifold of Calabi-Yau
manifolds
also get geometrized in M-theory on a manifold of $G_2$ holonomy.
This implies that large N dualities of \vaug\ lift to
{\it purely geometric transitions} of $G_2$ manifolds, with no
branes or flux.
It was shown in \refs{\amv\ach} that the large
$N$ duality of \vaug\ relating D6 branes on $T^*S^3$
and $O(-1)+ O(-1)\rightarrow \bf{P^1}$ with flux is
a geometric transition of a $G_2$ holonomy
manifold that is a spin bundle over $S^3$. For some other
work on manifolds of $G_2$ holonomy in the context of M-theory see
 \lref\achw{
B.~Acharya and E.~Witten,
``Chiral fermions from manifolds of G(2) holonomy,''
arXiv:hep-th/0109152.}
\lref\moh{J.A. Harvey and G. Moore, ``Superpotentials and
membrane instantons,'' [arXiv:hep-th/9907026].}
\lref\wan{
E.~Witten,
``Anomaly cancellation on G(2)manifolds,''
arXiv:hep-th/0108165.}
\lref\achc{
B.~S.~Acharya,
``Confining strings from G(2)-holonomy spacetimes,''
arXiv:hep-th/0101206.
}
\refs{\moh ,\hose,\kmcg,\gomis,\cvetic,\wan,\achw }.

In this section we consider the lift of type IIA compactifications
discussed in previous sections to $M$ theory.  We first construct
the $G_2$ manifold by asking what a transverse $M2$ brane probe
sees.  We then discuss what the
large $N$ dualities mean in terms of transitions in this context.

\subsec{M2 brane probes}

Consider a D2 brane probe of ${\cal M}_K$. In the absence of flux,
the theory on a single D2 brane is a three-dimensional ${\cal
N}=2$ gauge theory, and ${\cal M}_K$ arises as the classical Higgs
branch. If ${\cal M}_K$ is a toric manifold, the theory on the D2
brane at a generic point in ${\cal M}_K$ is, apart from the
``center of mass'' $U(1)_0$ gauge field, the linear sigma model
introduced in \phases\ .

Namely, the theory has gauge group $U(1)_0 \times G$
where $G =\prod_{a=1}^N U(1)_a$, and $N+3$ chiral multiplets
$X_i, i=1,\ldots N+3$ together with D-terms:
\eqn\Dt{\sum_i Q_a^i |X_i|^2 = r_a,}
where $Q_a^i, a=1,\ldots N$ gives the charge vector
under $U(1)_a$.
Then ${\cal M}_K$ is given by a set of minima of D-term potential, modulo
$G$. We used this well known construction in the previous sections.

Note that generators of the compact cohomology $H^{1,1}({\cal M}_K)$
are in one to one correspondence with the gauge-group factors.
Moreover, the periods of the complexified Kahler form
in $H_2({\cal M}_K,\bf{Z})$
are linear in Fayet-Illiopolous parameters $r_a$.
The theory on the D2 brane flows in the infra-red  $g^2 \rightarrow \infty$,
to a non-linear sigma model on ${\cal M}_K \times S^1$, where the $S^1$
factor comes from dualizing the c.o.m $U(1)_0$ to a periodic scalar $\phi
$. This theory is naturally interpeted as arising
on the M2 brane probe in M-theory compactification on ${\cal M}_K \times
S^1$.

Consider now turning on Ramond-Ramond two-form flux $F^{RR}$
whose class $[F^{RR}]\in H^2({\cal M}_K;\bf{Z})$ is
\eqn\flux{\int_{S^2_a}F^{RR} = N^a.}
Ramond-Ramond flux breaks supersymmetry of the bulk theory from
${\cal N}=2$ to ${\cal N}=1$. The RR flux reduces the number
of linearly realized supersymmetries in the world-volume theory of the
D-brane to two real supercharges, or ${\cal
N}=1$ in three dimensions.
Moreover, the flux must deform the D2 brane theory so that the Higgs
branch is topologically the $G_2$ manifold of the corresponding M-theory
compactification.

Let $A_0$ denote the gauge field of the
c.o.m. $U(1)$ and $A_a$ be the gauge fields of $G$.
Then, as we will argue below, the above flux turns on the coupling:
\eqn\BF{\sum_a N^a \int A_0 \wedge F_a,}
on the D2 brane world volume. In string theory, this
is completed to an interaction
that respects ${\cal N}=1$ supersymmetry, but breaks ${\cal N}=2$
\ref\susy{S.J. Gates, M.T. Grisaru, M. Rocek, W. Siegel, ``Superspace $or$
One Thousand and One Lessons in Supersymmetry'', Frontiers in Physics, 58,
The Benjamin/Cummings Publishing Co., 1983}.
We derive this in the general setting.  Later,
in a special case, we provide an alternative derivation
of the above result.

The instantons of the two dimensional gauge theory (which are
vortices in the three-dimensional sense) are field configurations that
describe holomorphic maps from the worldvolume into ${\cal M}_K$ \phases .
In particular in the instanton sector with wrapping numbers $n^a$
we have $\int F_a= k_a$.
Now we recall that
on the D2 brane world volume there is a Wess-Zumino term
\eqn\WZ{\int C^{RR} \wedge F_0 = \int F^{RR} \wedge A_0.}
For an instanton configuration, the $F^{RR}$ pulls back
$N^a k_a$ where $N^a$ is the corresponding
$RR$ flux through the $a$-th cycle.  However, this term
can be generated in the worldvolume theory
by the substitution
$F^{RR}\rightarrow N^a F_a$, noting
the relation of the instanton number
with the vortex number, which thus means that this is equivalent
to having the Chern-Simons term
$$\sum_a N^a \int F_a\wedge A_0$$
on the D2 brane worldvolume as claimed above.

For a special case
we can also derive the above result from a different point of view.
Let ${\cal M}_K =O(-1)+ O(-1)\rightarrow \bf{P^1}$,
with $N$ units of RR flux through the $\bf{P^1}$, and consider a D2 brane
on this space.
If we blow down the $\bf{P}^1$, the manifold develops a singularity
and the D2 brane theory develops a new branch where
the brane splits into a D4 brane and an anti-D4 brane wrapped
around the vanishing ${\bf P^1}$. In terms of the wrapped
D4 branes, the effect of the RR flux is transparent, and enters the action
via the WZ term
$ \int C^{RR}\wedge F\wedge F = \int_{S^2} F^{RR}\int A\wedge F$.
It remains to relate the gauge fields $A_{\pm}$ on the D4 brane
and the anti-D4 brane to those on the D2 brane linear sigma
model, and the answer
for this is (see e.g. \ref\klebwit{I.~R.~Klebanov and E.~Witten,
``Superconformal field theory on threebranes at a Calabi-Yau  singularity,''
Nucl.\ Phys.\ B {\bf 536}, 199 (1998),
[arXiv:hep-th/9807080].})
$$A_{\pm} = \frac{1}{2}(A_1\pm A_0),$$
where $A_0$ gauges the c.o.m. $U(1)_0$ as above, and
one recalls that $G=U(1)_1$. In terms of $A_{\pm}$,
the four chiral multiplets transform as ``bifundamentals''.
{}From the WZ terms we then have
$$N \int (A_+\wedge F_+ - A_-\wedge F_-) = N\int A_0 \wedge F_1,$$
where the left hand side takes into account opposite worldvolume
orientation of the D4 brane and the anti-D4 brane. This is in
agreement with the general result derived above.

\subsec{Higgs branch as a $G_2$ holonomy manifold}

We now want to understand how this Chern-Simons
term affects the Higgs branch, which gives the geometry
of the target space as seen by the M2 brane.
The answer is simple. In relating the D2 brane of type IIA to M2 brane
in M-theory, the c.o.m.
$U(1)_0$ gauge field $A_0$ is dual to the compact scalar $\phi$ on the M2 brane
parameterizing the position of the brane on the M-theory circle.
The only coupling of $A_0$ to the linear sigma model
 sector is via the interaction
\BF . To dualize $A_0$, introduce a dynamical vector field $B$
and replace the action
$\int| F_0|^2 + \sum_a \int A_0 \wedge N^a F_a$
with $\int |B|^2 + (B + \sum_a N^a A_a)\wedge F_0$.
Completing the square in $B$, the $B$ path integral becomes gaussian,
and integrating over $B$ the action reduces to previous one. On the other
hand, integrating over $A_0$ we find
$$d(B+\sum_a N^a A_a)=0 \rightarrow B = d\phi -\sum_a N^a A_a,$$
and the action becomes
\eqn\ta{\int |d\phi -\sum_a N^a A_a|^2,}
and the dual scalar $\phi$  picks up a logarithmic charge. Note that
$\phi$ is a real periodic variable.

The logarithmically charged real scalar does not affect the D-terms
as it is coupled minimally in the ${\cal N}=1$ sense \susy . The Higgs branch
is thus the space of minima of potential,
\eqn\Dt{\sum_i Q_a^i |X_i|^2 = r_a,}
in ${\bf C}^{N+3} \times S^1$ modulo $N$ gauge equivalences:
\eqn\gact{X_i \rightarrow X_i e^{i \sum_a Q^i_a  \lambda^a},
\quad \phi \rightarrow \phi + \sum_a N^a \lambda^a.}
This gives a manifold ${\cal G}$ which is guaranteed to admit
a metric of $G_2$ holonomy.

How do we recover the type IIA string theory we started with?
Since $\phi$ is identified as the 11-th circle, ignoring it
would give the type IIA geometry, which is indeed the Calabi-Yau
3-fold.  Moreover the fact that the $U(1)^a$'s act on $\phi$ in
the way indicated above, implies that, upon including $\phi$, the
corresponding $S^2_a$ becomes a base of a Hopf fibration
$S^3_a\rightarrow S^2_a$ with flux number $N^a$. This is indeed
the expected flux from the viewpoint of the type IIA theory.

We now apply this to the examples above:

\subsec{$G_2$ lift of ${\cal M}_K$ with $(h_{1,1},h_{2,1})=(2,0)$}
In this case, ${\cal M}_K$ is given by \egone\
$$|x_1|^2+|x_4|^2 - |x_2|^2-|x_5|^2 =s,\quad
                    |x_3|^2+|x_5|^2-2|x_4|^2 = t,$$
with two $U(1)$ actions of charges
$Q_s = (1,-1,0,1,-1),\;Q_t=(0,0,1,-2,1).$
We consider turning on $N$ units of RR flux through the $\bf{P^1}$
at $x_2=0=x_5$.

Following the discussion above, to get ${\cal G}$ we add the M-theory
circle scalar $\phi$ that has logarithmic charge
$N$ under $U(1)_s$.
Using $U(1)_s$ we can fix $\phi$ to arbitrary value, leaving an unbroken
${\bf Z}_N$ subgroup of the $U(1)$ under which $\phi$ is invariant.
As a consequence, ${\cal G}$ is an orbifold by  ${\bf{Z_N}}\in U(1)_s$.

Consider now the topology of the manifold in M-theory.
This has two compact three-cycles.

At $x_2=0=x_5$, the D-terms give
${|x_1|^2+|x_4|^2 =s,\;\;|x_3|^2=2|x_4|^2 +t.}$
For finite $t$, $x_3$ is determined in terms of $x_{1,2}$ and
it never vanishes, so $U(1)_t$ simply removes its
phase. We are simply left with the three-cycle
$$|x_1|^2+|x_4|^2 =s.$$
The ${\bf Z}_N$ action identifies
$$x_1, x_4 \sim x_1 e^{2 \pi i/N}, x_1 e^{2 \pi i/N},$$
so the three cycle is in fact $S^3/{\bf Z}_N$.

The other three-cycle is at $x_2=0=x_4$, where
${|x_3|^2+|x_5|^2 =t,\;\;|x_1|^2=|x_5|^2 +s.}$
Dividing by $U(1)_t$  together with the first equation gives an $S^2$.
{}From the second equation,
we can solve for $|x_1|^2$ but not for its phase, as
the discrete group action only makes the corresponding circle smaller.
This gives an $S^2\times S^1$.

We can compactify this to type IIA string theory by
``forgetting'' $\phi$, or equivalently introducing a new
$U(1)$ action under which $\phi$ has charge one and the rest
are neutral. This gives back ${\cal M}_K$
with $N$ units of flux through the $S^2$ which comes from the
$S^3/{\bf Z_N}$.

\subsec{$G_2$ lift of ${\cal M}_K$ with $(h_{1,1},h_{2,1})=(4,0)$}

Consider the Calabi-Yau manifold ${\cal M}_K$ given by the D-terms \egbb\
$$|x_0|^2+|x_1|^2 -|x_2|^2-|x_6|^2 = {s_{1}},
\quad |x_3|^2+|x_6|^2 -2|x_0|^2= {t_{1}},$$
$$|x_0|^2+|x_5|^2 -|x_4|^2-|x_6|^2 = {s_{2}},\quad
|x_2|^2+|x_4|^2 -2|x_0|^2= {t_{2}},$$
modulo $G=U(1)^4$. The corresponding charges of $x's$
can be read off from above:
$Q_{s_1}=(1,1,-1,0,0,0,-1)$, $Q_{t_1}=(-2,0,0,1,0,0,1)$
and $Q_{s_2}=(1,0,0,0,-1,1,-1)$, $Q_{t_2}=(-2,0,1,0,1,0,0).$

Let us turn on $N_1$ units of flux through the ${\bf P^1}$
at $x_2=0=x_6$, and  $N_2$ units of flux through the ${\bf P^1}$
at $x_4=0=x_6$.
The M-theory manifold ${\cal G}$ has an additional periodic scalar
$\phi$ which has charge $N_1$ under the first and $N_2$ under the third
$U(1)$ and zero for others.

The geometry of the minimal cycles is as follows.
At $x_2=0=x_6$, we find
\eqn\stta{|x_1|^2+|x_0|^2 =s}
and $|x_3|^2=2 |x_0|^2 +{t_1}, |x_4|^2=2|x_0|^2 +{t_2}$,
and $|x_5|^2=|x_0|^2 +{t_1}+{s_2}.$
For finite $t$, $|x|^2_{3,4,5}$ are determined in terms of $x_{0,1}$
and never vanish. We can use $U(1)_{s_2, t_1,t_2}$ to
fix the phases of $x_{3,4,5}$
This leaves \stta\ together with $U(1)_{s_1}$ action:
$$x_{0,1}\rightarrow e^{i \lambda} x_{0,1},\quad \phi \rightarrow
\phi+ N_1 \lambda,$$
which is an $S^3/{\bf Z}_{N_1}$.

Considering the locus $x_4=0=x_6$, analogous argument
gives an $S^{3}/{\bf Z}_{N_2}$.

\subsec{$G_2$ transitions} Up to now we have talked about how to
lift the type IIA geometries we have considered in the phase where
we have $S^2$'s with fluxes through them.  For each one of them,
we can have two other phases in the type IIA description:  One
where the $S^2$ undergoes a flop to another $S^2$ with the
corresponding Kahler parameter $t\rightarrow -t$ and another
transition where $S^2\rightarrow S^3$.  The transition of the flop
type already can be described as we have done above and lifts in
M-theory to smooth $G_2$ manifolds.  The other transition, as we
know from the type IIA, involves D6 branes wrapped around $S^3$'s,
which would lift, in M-theory, to singular $G_2$ geometries where
locally each corresponding $S^3$ is the locus of an $A_{N_i-1}$
singularity.  Aspects of this local three phases system has been
studied from the viewpoint of M-theory on $G_2$ holonomy manifold
in \awi .  An equivalent setup where the three phase system can be
seen and studied perturbatively in string theory was discussed in
\AVG .

In the examples we have studied the local geometries are of the
kind already studied, thus the same transition still continues to
be valid, except that sometime the corresponding $S^3$'s are not
independent, and represent the same class in the $G_2$ holonomy
manifold.  This would be the case for our second example
where the corresponding $S^3$'s are in the same class, one would
be a locus of an $A_{N_1-1}$ singularity and the other would be a
locus of an $A_{N_2-1}$ singularity, as would follow from the lift
of the corresponding type IIA geometry.  Note that for the case of
$N_1=N_2=1$ the lift of both sides to M-theory involves smooth
$G_2$ holonomy manifolds.  In that case the transition
$(h_{1,1},h_{2,1})=(4,0)\rightarrow (2,1)$ implies a transition of
local $G_2$ holonomy manifolds. On the side corresponding to the
lift of $(h_{1,1},h_{2,1})=(4,0)$ we have the local $G_2$ manifold
with compact betti numbers given by $(b_2,b_3,b_4,b_5)= (2,4,1,1)$
where the 2-cycles can be realized by two $S^2$'s (corresponding
to the $t_i$) the four 3-cycles are realized by two $S^3$'s
corresponding to $s_i$'s and two $S^2\times S^1$'s corresponding
to $t_i$, the $b_4$ corresponds to $S^2\times S^2$ realized by the
$t_1,t_2$ spheres, and the $b_5$ corresponds to the same 4-cycle
times the 11-th circle.  On the side with
$(h_{1,1},h_{2,1})=(2,1)$ we have the local $G_2$ manifold with
compact betti numbers given by $(b_2,b_3,b_4,b_5)=(2,3,0,0)$ as
one can verify by considering the lift of various cycles to
M-theory. It is amusing to note that with our techniques, not only
we are predicting a smooth transition among these $G_2$ holonomy
manifolds, but we are actually computing the exact superpotential
generated by the M2 branes wrapping the non-trivial 3-cycles,
which is the lift of the superpotential computations we have done
here, as noted in similar cases in \AKV.

\newsec{Linear sigma models, $G_2$ manifolds and mirror symmetry}

In the previous section we found the worldsheet theory seen
by an M2 brane propagating on a $G_2$ holonomy manifold.  If we
compactify M-theory on a further circle down to three spacetime
dimensions, we can identify the worldvolume theory on M2 brane wrapping
the $S^1$ with the worldsheet probe in the same $G_2$ holonomy manifold.
This can be viewed as an ${\cal N}=1$ deformation of an ${\cal N}=2$
supersymmetric gauge system as discussed before.  Namely we have
the $\prod_{a=1}^N U(1)_a$ gauge system
with $X_i$ chiral superfields, where
$i=1,...,N+3$,
with charges $Q_a^i$, and a real periodic scalar field $\phi$ with
shift symmetry gauged according to
$$\phi \rightarrow \phi +\sum_a \lambda_a N_a$$
Note that $\phi$ is an ${\cal N}=1$ multiplet and it only couples
to the ${\cal N}=1$ gauge supermultiplets of the ${\cal N}=2$
system.
 This coupling breaks the ${\cal N}=2$ supersymmetry
of the worldsheet to ${\cal N}=1$.
\lref\achmir{B.S. Acharya, ``On Mirror Symmetry for Manifolds of Exceptional
Holonomy,'' Nucl.Phys. {\bf B524} 269 (1998), [arXiv: hep-th/9707186].}
Note that the vanishing of the $\beta$- function  for the FI-term that
was crucial for existence of a conformal fixed point in the IR for
the Calabi-Yau linear sigma model is not spoiled, perturbatively, by the
addition of the periodic scalar $\phi$ with logarithmic charge. 
This suggests that there is no obstruction in the present case for this
theory to flow to a conformal theory corresponding to a sigma model on a
$G_2$ holonomy manifold.

It is natural to apply the idea of mirror symmetry to this
context. For previous studies of mirror symmetry in the context of
$G_2$ holonomy manifolds see \refs{\achmir ,\moh}. The idea is
rather simple: we dualize the phases of the linear sigma model
fields $X_i$, as in \HV, as well as the circle represented by
$\phi$. Let us first discuss the dualization of the circle $\phi$.
Recall that in the D2 brane worldvolume this was dual to a
$U(1)_0$ gauge field which coupled to the other gauge fields by a
BF Chern-Simons terms.  Dualizing the 2d scalar $\phi$ in the 2d
sense, is equivalent to first dualizing $\phi$ in the 3d theory in
terms of gauge field, and then identifying the dual scalar
$\theta$ with the component of the gauge field along the circle
taking us from three dimensions to two dimensions 
\lref\HoriAX{
K.~Hori and A.~Kapustin,
``Duality of the fermionic 2d black hole and N = 2 Liouville theory as  mirror symmetry,''
JHEP {\bf 0108}, 045 (2001)
[arXiv:hep-th/0104202].}\lref\AganagicUW{
M.~Aganagic, K.~Hori, A.~Karch and D.~Tong, ``Mirror symmetry in
2+1 and 1+1 dimensions,'' JHEP {\bf 0107}, 022 (2001)
[arXiv:hep-th/0105075].}\refs{\HoriAX,\AganagicUW}.  
Note that this gives rise in the 2d
theory to an interaction of $\theta$ with the ${\cal N}=2$ gauge
system  by
$$\sum_{i=1}^N N_i\int d^3x \ A_0 F_i\rightarrow \sum_i N_i\int d^2x
\theta F_i$$
In other words, for a fixed value of $\theta$ this effectively
shifts the theta angle of the $i$-th gauge group by $N_i\theta$.

If we stop at this stage we have a dual description of the $G_2$
holonomy manifold, in terms of a Calabi-Yau 3-fold times a circle
parameterized by $\theta$ where the fiber Calabi-Yau has some
$B$-field moduli varying over the circle. More precisely, the
B-field moduli are the theta angles in the gauge theory, and so in
this description we have $H_{NS}$ flux. This kind of mirror
symmetry exchanges the type IIA and type IIB superstrings because
it involves the application of a single T-duality.

It is more interesting
to get a purely geometric description for the mirror
without a flux.  For this purpose,
we can continue the dualization to the rest of the fields
using the results of \HV .
  For this purpose, we assume, using the adiabatic principle,
that we have a slowly varying $\theta $ field. In this way we obtain
a dual geometry, as in \HV\ but for which the $\theta$ variable appears
in the defining equation.  More precisely, it was shown in \HV\ (see
also \hiv ) that the mirror geometry for the local 3-fold is given by
$$xz=F(u,v,t_i)$$
where $x,z,u,v$ are complex coordinates and $t_i$ are the
complexified Kahler parameters, which on the mirror play the role
of complex structure. Here, all we have to do to get the $G_2$
mirror is to replace $t_i\rightarrow t_i +iN_i\theta$.  In other
words the $G_2$ mirror is given by the locus in ${\bf
C^2}\times{\bf C^2}^*\times {\bf S^1}$ satisfying
$$xz=F(u,v,t_i+iN_i\theta)$$
Note that this mirror symmetry takes type IIA strings back to type IIA
strings, and type IIB strings back to type IIB strings, as we have dualized
the extra circle $\phi$ in addition to that of the Calabi-Yau 3-fold.
It would be interesting to see what new predictions this mirror formulation
would lead to in the context of compactifications of superstrings on
$G_2$ holonomy manifolds.

\centerline{\bf Acknowledgements}
We would like to thank F. Cachazo, S. Gukov,
K. Hori, M. Marino and M. Wijnholt for valuable discussions.

This research is supported in part by NSF grants PHY-9802709
and DMS 0074329.
\listrefs
\end